\documentclass{elsart}

\usepackage{graphics}
\usepackage{amssymb}
\usepackage{epsfig}
\usepackage[english]{babel}

\newtheorem{e-proposition}[theorem]{Proposition}

\newtheorem{e-definition}[theorem]{Definition\rm}

\renewcommand{\theequation}{\arabic{equation}}
\begin{document}

\begin{frontmatter}

\selectlanguage{english}
\title{\protect\textbf{Conventional Magnetic Superconductors: Coexistence of Singlet Superconductivity and Magnetic Order}}

\author{Miodrag L. Kuli\'{c} $^{1,2}$},
\address{\protect{$^{1}$}Institute for Theoretical Physics, J. W. Goethe University, Frankfurt/Main, P.O.Box 111932 Frankfurt/Main, Germany\\ \protect{$^{2}$}CPMOH, Universit\'{e} Bordeaux and CNRS, 33405 Talence Cedex, France}

\medskip
\begin{center}
\end{center}

\begin{abstract}
The basic physics of bulk magnetic superconductors (MS) related to
the problem of the coexistence of singlet superconductivity (SC)
and magnetic order is reviewed. The interplay between exchange
(EX) and electromagnetic (EM) interaction is discussed and argued
that the singlet SC and uniform ferromagnetic (F) order
practically never coexist. In case of their mutual coexistence the
F order is modified into a domain-like or spiral structure
depending on magnetic anisotropy. It turns out that this situation
is realized in several superconductors such as $ErRh_{4}B_{4}$,
$HoMo_{6}S_{8}$, $HoMo_{6}Se_{8}$ with electronic and in
$AuIn_{2}$ with nuclear magnetic order. The later problem is also
discussed here.

The coexistence of SC and antiferromagnetism is more favorable
than with the modified F order. Very interesting physics is
realized in systems with SC and weak-ferromagnetism which results
in an very reach phase diagram.

The properties of magnetic superconductors in magnetic field are
very peculiar, especially near the (ferro)magnetic transition
temperature where the upper critical field becomes smaller than
the thermodynamical critical field.

The extremely interesting physics of Josephson junctions based on
MS with spiral magnetic order is also discussed. The existence of
the triplet pairing amplitude $F_{\uparrow \uparrow }$
($F_{\downarrow \downarrow }$) in MS with rotating magnetization
(the effect recently rediscovered in SFS junctions) gives rise to
the so called $\pi $-contact. Furthermore, the interplay of the
superconducting and magnetic phase in such a contact renders
possibilities for a new type of coupled Josephson-qubits.

\vskip0.5 
\baselineskip 
\end{abstract}

\end{frontmatter}

\selectlanguage{english} 

\section{\ Introduction}

The physics of magnetic superconductors is very interesting subject due to
the pronounced competition of magnetic order and singlet superconductivity
in bulk materials. The question of their coexistence was first rised
theoretically in the pioneering work by V. L. Ginzburg \cite{Ginzburg} in
1956, where only the \textit{electromagnetic (EM) interaction} between
magnetic moments and superconductivity was considered. However, the
breakthrough in the physics of MS came with experiments after the discovery
of \textit{ternary rare earth} ($RE$) compounds such as borides $%
(RE)T_{4}B_{4}$ with transition elements $T=Rh,Ir$, chalcogenides $%
(RE)Mo_{6}X_{8}$ ($X=S,Se$), silicides $(RE)_{2}T_{2}$Si$_{5}$ and stannides
$(RE)T_{x}Sn_{5}$ \cite{Maple}. In most of them type-II superconductivity is
realized and in all of them are the localized RE magnetic moments \textit{%
regularly} \textit{distributed} in the crystal lattice. The basic
crystallographic structure, for instance in RERh$_{4}$B$_{4}$, contains
localized moments (LMs) which are rather far a way from the Rh and B blocks
which deliver conduction electrons. Due to this spatial separation the
conduction electrons jump rarely on magnetic ions making the direct exchange
interaction (EX) $J_{sf}(\ll 10^{3}$ $K)$ much smaller than in transition
metallic magnets. In these systems the 4f rare-earth shells are responsible
for localized moments in which the f-level lies much below the Fermi energy,
$E_{f}<<E_{F}$. A number of compounds belonging to the above families have
shown coexistence of superconductivity with the antiferromagnetic (AF) order
- \textit{antiferromagnetic superconductors} (AFS), such as $(RE)Rh_{4}B_{4}$
(RE=Dy, Sm,...), and in most of them the Neel (AF) transition temperature $%
T_{N}$ is smaller than the superconducting one $T_{c}$, i.e. $T_{N}<<T_{c}$.

However, a lot of research, both experimental and theoretical, starting from
the late seventies were devoted to MS systems in which ferromagnetic (F)
order and singlet SC compete due to their strong antagonistic characters -
we call these systems \textit{ferromagnetic superconductors} ($FS$) and they
are the main subject of this review. It turned out that the modified F and
SC, order can under certain conditions coexist since the F order is
transformed (in the presence of superconductivity) into a spiral or
domain-like structure - depending on the type and strength of magnetic
anisotropy in the system \cite{BuBuKuPaAdvances}, \cite{BuKuRu}. In the RE
ternary compounds this competition is rather strong and therefore these two
orderings coexist in a limited temperature interval $T_{c2}<T<T_{m}$ (the
reentrant behavior), for instance in ErRh$_{4}$B$_{4}$ and HoMo$_{6}$S$_{8}$%
. The coexistence region in ErRh$_{4}$B$_{4}$ is narrow with $T_{m}\approx
0.8$ $K$, $T_{c2}\approx 0.7$ $K$ and $T_{c}=8.7$ $K$, while for HoMo$_{6}$S$%
_{8} $ it is even narrower with $T_{m}\approx 0.74$ $K$, $T_{c2}\approx 0.7$
$K$ and $T_{c}=1.8$ $K$ - see Refs. \cite{Maple}, \cite{BuBuKuPaAdvances},
\cite{BuKuRu}. In HoMo$_{6}$Se$_{8}$ where $T_{c}=5.5$ $K$, $T_{m}\approx
0.8 $ the exchange interaction is weaker and the coexistence persists down
to $T=0$ $K$.

A new and very interesting research field in the physics of ferromagnetic
superconductors was opened in 1997 by Pobell's group in Bayreuth \cite%
{Pobell}, which discovered the coexistence of \textit{superconductivity and
nuclear magnetic order} in the type-I superconductor $AuIn_{2}$ with $%
T_{c}=0,207$ $K$ and $T_{m}=35$ $\mu K$. Although there is a tendency to the
nuclear ferromagnetic order, superconducting electrons enforce the
appearance of a spiral or domain-like nuclear magnetic ordering in the SC
state\ below $T_{m}$ \cite{KuBuBu} .

It turns out that not only bulk properties of FS are exotic, but also
Josephson junctions made of bulk MS with spiral ordering show potentially
fascinating properties, such as $\pi $-contact \cite{Kic-Ig}, combination of
a magnetic analog of the Josephson effect for spin current with the ordinary
Josephson effect for charge current \cite{Kic-Ig-QC}.

In the following we shall discuss mainly the \textit{microscopic and
macroscopic theory of ferromagnetic and antiferromagnetic superconductors}
which takes into account the most relevant interactions between localized
moments and conduction electrons - the \textit{exchange} (EX) and \textit{%
electromagnetic} (EM) interaction. This theory was elaborated by Buzdin,
Bulaevskii, Panyukov and present author - see \cite{BuBuKuPaAdvances} and
references therein, and successfully applied to a number of systems. Due to
the lack of space we shall discuss effects of magnetic field on magnetic
superconductors briefly - for this \ subject we refer the reader to Ref.\cite%
{BuBuKuPaAdvances}.

We would like to point out here that in the last several years there was a
huge activity in studying of hybrid heterogeneous magnetic superconductors
such as S-F multilayers and S-F-S Josephson junctions. This field is not
only of importance for the fundamental solid state physics but it is of
enormous interest for applications in spintronics and quantum computing,
especially after the experimental confirmation of the remarkable prediction
of the $\pi $-Josephson contact by Alexander Buzdin and coworkers \cite%
{Buzdin-pi-shift}, \cite{Buzdin-Review}. This exciting field will be covered
elsewhere in this issue, as well as the physics of other magnetic
superconductors - heavy fermions, borocarbides $(RE)Ni_{2}B_{2}C$, cuprates
RuSr$_{2}$GdCu$_{2}$O$_{8}$, ferromagnets with triplet SC such as UGe$_{2}$.

\section{Competition between SC and F order in FS}

Here we shall be limited to those magnetic superconductors where the
magnetic ordering of the \textit{localized 4f moments }(LM)\ is due to the
indirect exchange interaction (RKKY) going via the conduction electrons. The
characteristic exchange energy is $\theta _{ex}\approx N(0)h_{0}^{2}$, where
$N(0)$ is the density of states at the Fermi level (per LM) and $%
h_{0}(=(g-1)nJ_{sf}(0)\langle \hat{J}_{z}\rangle )$ is the maximal exchange
field acting on conduction electrons. Here, g is the Lande factor, n is the
density of localized magnetic moments (LMs), $J_{sf}(0)$ is the direct
exchange energy between conduction electrons and LMs, $\langle \hat{J}%
_{z}\rangle $ is the averaged total angular moment of the LM. Note that the
exchange field acting on electrons is $h_{ex}(\mathbf{r})=h_{0}S(\mathbf{r})$%
, where $S(\mathbf{r})(=\langle \hat{J}_{z}\rangle /J)$ is the localized
spin normalized to one. Let us mention in advance that in the RE ternary
compounds the exchange field $h_{0}$ is still rather large, i.e $h_{0}\sim
10^{2}$ $K$ and $h_{0}\gg \Delta _{0}\lesssim 10$ $K$. We shall see below
that in spite of the fact that $h_{0}$ is larger than the Clogston
paramagnetic field $h_{p}$, i.e. $h_{0}\gg h_{p}\approx 0.7\Delta _{0}$,
there is a coexistence of SC and modified ferromagnetic order. In the
presence of magnetic ordering characterized by the magnetization $\mathbf{M}(%
\mathbf{r})$ there is electromagnetic interaction between localized moments
and (super)conducting electrons, since $\mathbf{M}(\mathbf{r})=n\mu \mathbf{S%
}(\mathbf{r})$ creates the dipolar magnetic field $\mathbf{B}(\mathbf{r})=rot%
\mathbf{A}(\mathbf{r})$ which on the other side induce screening current $%
\mathbf{j}_{s}$ of conduction electrons (the Meissner effect). The Fourier
transformed $\mathbf{j}_{s}$ is related to $\mathbf{A}$ by the kernel $K_{s}(%
\mathbf{q})$, i.e. $\mathbf{j}_{s}(\mathbf{q})=-K_{s}(\mathbf{q})\mathbf{A}(%
\mathbf{q})$. Having in mind magnetic superconductors based on RE ternary
compounds we shall discuss the physics in the mean-field approximation for
SC and magnetic subsystem - quite appropriate approach, i.e. we put $\mathbf{%
\hat{S}}(\mathbf{r})\rightarrow \mathbf{S}(\mathbf{r})=<\mathbf{\hat{S}}(%
\mathbf{r})>$

The total Hamiltonian of the system is given by

\begin{equation}
\hat{H}=\hat{H}_{BCS}+\hat{H}_{CF}+\hat{H}_{imp}+\sum_{i}[-\mathbf{B}(%
\mathbf{r}_{i})g\mu _{B}\mathbf{\hat{J}}_{i}+\hat{H}_{CF}(\mathbf{\hat{J}}%
_{i})]  \label{1}
\end{equation}

\[
\hat{H}_{BCS}=\int d^{3}r\{\hat{\psi}^{\dagger }(\mathbf{r})[\hat{\epsilon}(%
\mathbf{\hat{p}-}\frac{e}{c}\mathbf{A})]-\mu ]\hat{\psi}(\mathbf{r})+\hat{%
\psi}^{\dagger }(\mathbf{r})\hat{V}_{ex}(\mathbf{r})\hat{\psi}(\mathbf{r})
\]

\begin{equation}
+\frac{1}{2}\Delta (\mathbf{r})\hat{\psi}^{\dagger }(\mathbf{r})i\sigma _{y}%
\hat{\psi}^{\dagger }(\mathbf{r})-\frac{1}{2}\Delta ^{\ast }(\mathbf{r})\hat{%
\psi}(\mathbf{r})i\sigma _{y}\hat{\psi}(\mathbf{r})+\frac{\mid \Delta (%
\mathbf{r})\mid ^{2}}{g_{epi}}\}.  \label{2}
\end{equation}%
Here, $\hat{\epsilon}(\mathbf{\hat{p}-}\frac{e}{c}\mathbf{A})$ is the band
energy of electrons in magnetic field, $\Delta (\mathbf{r})$ is the \textit{%
singlet} \textit{superconducting order parameter}, $\mathbf{\sigma =\{\sigma
}_{x}\mathbf{,\sigma }_{y}\mathbf{,\sigma }_{z}\mathbf{\}}$ are Pauli
matrices, while the \textit{exchange field} acting on electrons is given by

\begin{equation}
\hat{V}_{EX}(\mathbf{r})\mathbf{=}\left(
\begin{array}{cc}
h_{ex}^{z}(\mathbf{r}) & h_{ex}^{\perp }(\mathbf{r}) \\
h_{ex}^{\perp ,\ast }(\mathbf{r}) & -h_{ex}^{z}(\mathbf{r})%
\end{array}%
\right) .  \label{3}
\end{equation}%
We go slightly in advance by informing the reader, that in the SC state the
ferromagnetic order is modified into a spiral or domain-like structure with
the wave vector $\mathbf{Q}$, depending on magnetic anisotropy described by $%
\hat{H}_{CF}$. If the magnetic anisotropy is small (or easy plane) than the
spiral structure is realized with $h^{\perp }(\mathbf{r})=he^{iQz}\ $and $%
h^{z}(\mathbf{r})=0$, while in the opposite case with an easy axis
anisotropy one has $h^{\perp }(\mathbf{r})=0$ and $h^{z}(\mathbf{r})=h^{z}(%
\mathbf{r+L})$, $L=2\pi /Q$. The effect of nonmagnetic impurities is
described by $\hat{H}_{imp}$ whose effect is characterized by the mean-free
path $l$ and time $\tau $.

\subsection{Sinus magnetic structure due to SC for $T\lesssim T_{m}$}

In the RE ternary magnetic superconductors in which the singlet SC and
ferromagnetic order compete, the superconducting critical temperature, $%
T_{c} $, is much higher than the magnetic one, i.e. $T_{m}<<T_{c}$. Before
discussing the complete phase diagram we shall study the coexistence problem
at temperatures near $T_{m} $, i.e. $T\lesssim T_{m}$, where the magnetic
order parameter is small $S<<1$. In case when the easy-axis magnetic
anisotropy $D$ is sufficiently large then the sinus structure $\mathbf{S}(%
\mathbf{r})\approx S(T)\sin \mathbf{Qr}$ appears below $T_{m}$ (for small $D$
a spiral order is favored - see \textit{2.3}). In that case $h_{ex}(\mathbf{r%
}) =h_{0}\mid \mathbf{S}(\mathbf{r})\mid <<h_{0},\Delta $ and the
free-energy can be calculated by the perturbation theory

\begin{equation}
F\{\mathbf{S}(\mathbf{r}),\Delta (\mathbf{r}),\mathbf{A}(\mathbf{r}%
)\}=F_{M}\{\mathbf{S}\}+F_{S}\{\Delta \}+F_{Int}\{\mathbf{S},\Delta ,\mathbf{%
A}\},  \label{4}
\end{equation}%
where $F_{M}$ and $F_{S}$ are the magnetic and SC functional without mutual
interaction,respectively.

\[
F_{M}\{\mathbf{S}(\mathbf{r})\}=n\sum_{q}\{\frac{1}{2}[(T-T_{m0})+\theta
a^{2}q^{2})]\mid \mathbf{S}(\mathbf{q})\mid ^{2}-D\mid S_{z}(\mathbf{q})\mid
^{2}\}
\]

\begin{equation}
+\int d^{3}r\frac{(\mathbf{B}-4\pi \mathbf{M})^{2}}{8\pi },  \label{5a}
\end{equation}%
where $\vec{S}(\vec{q})$ is the Fourier transform of $\vec{S}(\vec{r})$. The
last term in Eq. (5) is the magnetic energy for a given magnetization $%
\mathbf{M}(\mathbf{r})=n\mu \mathbf{S}(\mathbf{r})$ and the magnetic
induction $\mathbf{B}=4\pi \mathbf{M}$. The characteristic energy for the EM
interaction is given by $\theta _{em}=(B^{2}/8\pi n)=2\pi n\mu ^{2}$ which
is $\sim 1$ $K$ in the RE ternary compounds.

Since $T_{m}<<T_{c}$ and $h\ll \Delta $ the SC free-energy density ($%
F_{S}=\int d^{3}r\tilde{F}_{S}$)

\begin{equation}
\tilde{F}_{S}\{\Delta (\mathbf{r})\}=-\frac{1}{2}N(0)\Delta ^{2}\ln \frac{%
e\Delta _{0}^{2}}{\Delta ^{2}}  \label{5}
\end{equation}%
is minimized for $\Delta \approx \Delta _{0}$ and we omit it from the
analysis near $T_{m}$. The part $F_{Int}$ describes the EX and EM
interaction between SC and magnetic order (note $\mathbf{j}_{s}(\mathbf{q}%
)=-K_{s}(\mathbf{q})\mathbf{A}(\mathbf{q})$)

\begin{equation}
F_{Int}=\sum_{q}\{\frac{1}{2}K_{s}(\mathbf{q})\mid \mathbf{A}(\mathbf{q}%
)\mid ^{2}+\theta _{ex}\frac{\chi _{n}(\mathbf{q})-\chi _{s}(\mathbf{q})}{%
\chi _{n}(0)}\mid \mathbf{S}(\mathbf{q})\mid ^{2}\}.  \label{6}
\end{equation}

After minimizing $F\{\mathbf{S}(\mathbf{r}),\Delta (\mathbf{r}),\mathbf{A}(%
\mathbf{r})\}$ with respect to $\mathbf{A}(\mathbf{r})$ one gets $F\{\mathbf{%
S}(\mathbf{r}),\Delta (\mathbf{r})\}$ in the following form (see more below
and in \cite{BuBuKuPaAdvances})
\[
F\{\mathbf{S}(\mathbf{r}),\Delta (\mathbf{r})\}=n\sum_{q}\{\frac{1}{2}%
[(T-T_{m0})+\theta a^{2}q^{2})]\mid \mathbf{S}(\mathbf{q})\mid
^{2}-D_{z}\mid S_{z}(\mathbf{q})\mid ^{2}
\]

\begin{equation}
+\frac{\theta _{ex}}{2}\frac{\chi _{n}(\mathbf{q})-\chi _{s}(\mathbf{q})}{%
\chi _{n}(0)}\mid \mathbf{S}(\mathbf{q})\mid ^{2}+\frac{\theta _{em}}{2}%
\frac{4\pi K_{s}(q)\mid \mathbf{S}(\mathbf{q})\mid ^{2}+(\mathbf{qS}(\mathbf{%
q}))(\mathbf{qS}(-\mathbf{q}))}{q^{2}+4\pi K_{s}(q)}\}  \label{7}
\end{equation}%
Here, $a$ is of the order of lattice constant (magnetic stiffness) and the
bare critical temperature $T_{m0}$ and $\theta $ take in a subtle way (note $%
T_{m0}\neq \theta $ - see details in \cite{BuBuKuPaAdvances}) into account
the indirect EX and direct dipole-dipole (EM) interaction between LMs - see
\cite{BuBuKuPaAdvances}. $\chi _{n}(\mathbf{q})$ and $\chi _{s}(\mathbf{q})$
are electronic susceptibilities in the normal and SC state, respectively. $%
\theta _{em}=2\pi \mu ^{2}$ characterizes the EM effects in the
dipole-dipole interaction between LMs, while $K_{s}(q)$ is the EM kernel
which describes the screening effect of the dipole-dipole interaction by
thesuperconducting electrons. $D_{z}(>0)$ is the magnetic anisotropy which
orients spins along the z-axis.

Due to the singlet SC pairing $\chi _{s}(\mathbf{q})$ is reduced
significantly at small wave vectors $q<\xi _{0}^{-1}$ where $\xi _{0}$ is
the SC coherence length. In the \textit{clean limit} ($l\rightarrow \infty )$
and at $T=0$ one has $\chi _{s}(\mathbf{0})=0$ which means that the \textit{%
ferromagnetic order can not coexist with singlet superconductivity}. In $%
Fig.1$ we show $\chi _{s,n}(\mathbf{q})$ schematically for the cases when
the ferromagnetic (1a) or antiferromagnetic order (1b) is realized in the
normal state. It is seen that a singlet superconductor behaves as a normal
metal at large momenta, i.e. $\chi _{s}({q \sim k_{F}})$ is weakly affected
by SC. Therefore AF competes with SC much less than the ferromagnetic order
does.

We stress that at finite temperature $\chi _{s}(\mathbf{0})\neq 0$ but
exponentially small in singlet s-wave SC, while in d-wave SC one has $\chi
_{s}(\mathbf{0})\sim \chi _{n}(\mathbf{0})(T/\Delta _{0})$. In the presence
of the spin-orbit (SO) scattering $\chi _{s}(\mathbf{0})$ is also finite.
The general expression for $\chi _{s}(\mathbf{q})$ is calculated in \cite%
{Entin}

\begin{equation}
\chi _{s}(\mathbf{q})=1-\pi T\sum_{\omega _{n}}\frac{1}{(1+u_{\omega
}^{2})(P(\omega ,q)-1/2\tau _{1})},  \label{8a}
\end{equation}%
where $u_{\omega }=\omega /\Delta $ and

\bigskip
\begin{equation}
P(\omega ,q)=\frac{1}{2}\frac{qv_{F}}{\arctan \{\frac{qv_{F}}{2\Delta }\sqrt{%
1+u_{\omega }^{2}+1/2\tau _{-}}\}}.  \label{8b}
\end{equation}%
Here, $\tau _{1}^{-1}=\tau _{-}^{-1}-(4/3)\tau _{so}^{-1}$, $\tau
_{-}^{-1}=\tau ^{-1}+\tau _{so}^{-1}$, $l_{so}=v_{F}\tau _{so}$ and $\omega
_{n}=\pi T(2n+1)$. Later, we shall discuss some effects of the SO
interaction on the coexistence phase. The effect of the exchange scattering
is similar, i.e. $\chi _{s}(\mathbf{0})$ is finite for finite $\tau _{s}$.

Since in the following we study the competition between SC and
ferromagnetism at low temperatures it is sufficient to give a general
expression for $K_{s}(q)$ in the clean limit

\begin{equation}
K_{s}(q)=\frac{3n_{e}\Delta }{mcqv_{F}}\int_{0}^{1}dx\frac{1-x^{2}}{x}\frac{%
arcsh(\frac{xqv_{F}}{2\Delta })}{\sqrt{1+(\frac{xqv_{F}}{2\Delta })^{2}}}.
\label{8c}
\end{equation}
The expression for finite $l$ is cumbersome and is omitted here. Some
limiting cases of $K_{s}(q)$ which are relevant for real magnetic
superconductors will be studied below.
\begin{figure}[tbp]
\resizebox{.5 \textwidth}{!} {
\includegraphics*[width=7cm]{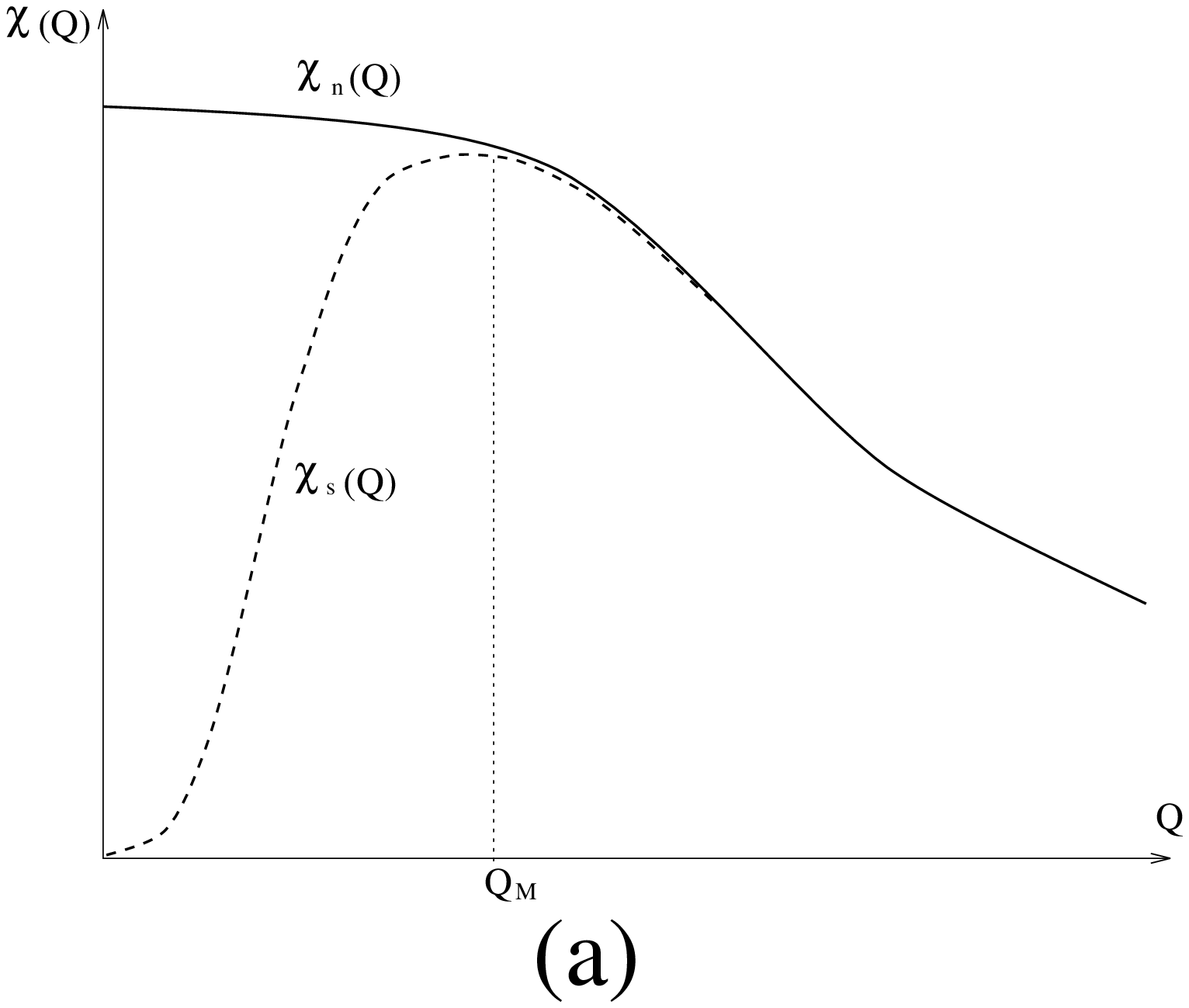}} \includegraphics*[width=7cm]{%
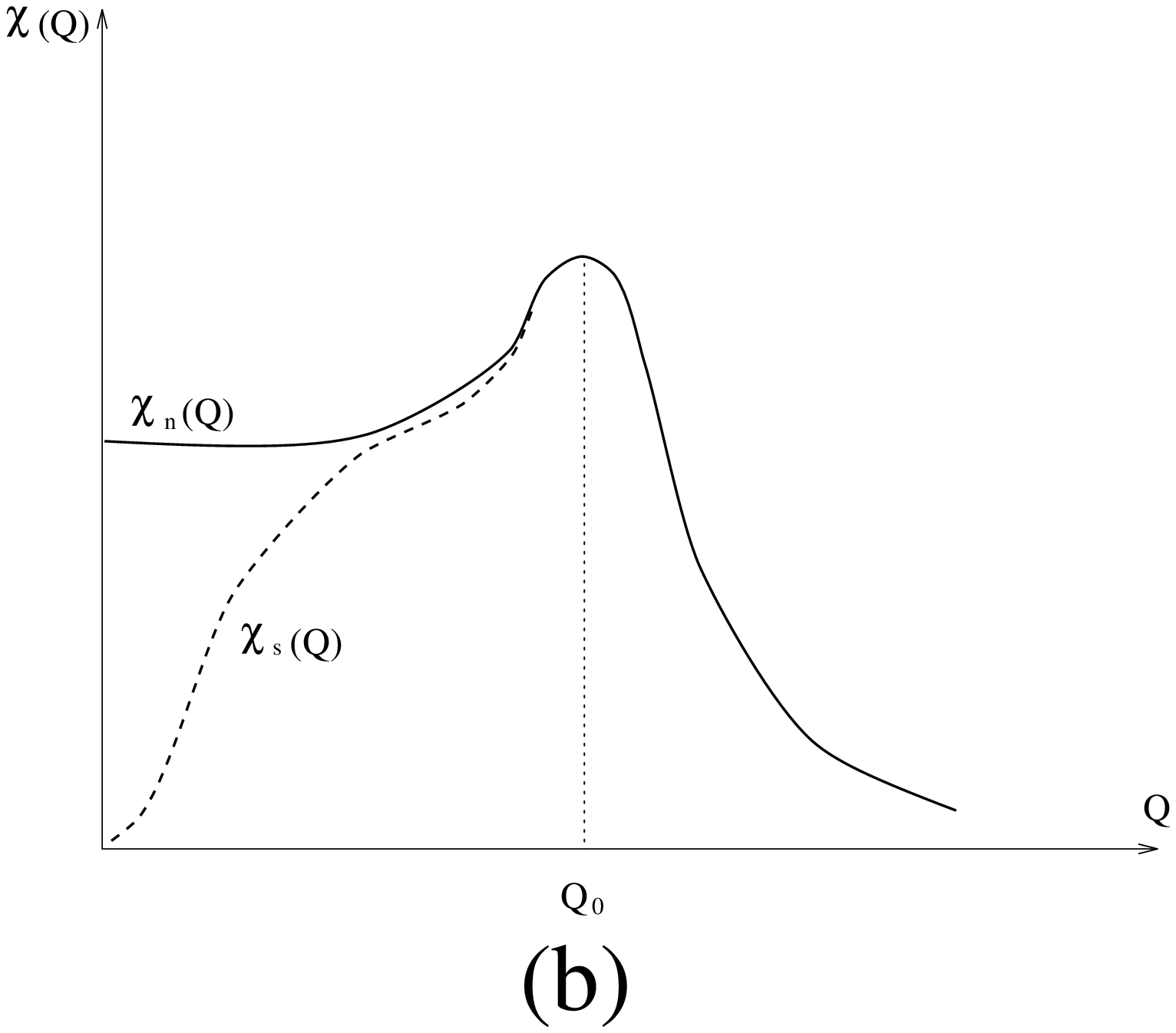}
\caption{Schematic spin susceptibility in the SC and normal state $\protect%
\chi _{n,s}(\mathbf{q})$: \textbf{(a)} for the ferromagnetic order in the
normal state peak at $Q=0$; \textbf{(b)} for the antiferromagnetic order -
peak at $Q_{0}$.}
\end{figure}

By minimizing $F\{\mathbf{S}(\mathbf{r}),\Delta (\mathbf{r})\}$ w.r.t. the
wave vector $q$ one obtains the equilibrium magnetic structure which depends
on microscopic parameters $a,\xi _{0},\lambda _{L}$(the London penetration
depth), $\theta _{ex},\theta _{em}$. From the above equation we conclude
that due to the EM interaction the magnetic structure is \textit{transverse}%
, $\mathbf{q\cdot S}(\mathbf{q})=0$.

Let us analyze $\chi _{s}(\mathbf{q})$ and $K_{s}(q)$ in the interesting
range of parameters. In the \textit{clean limit} and for \ $q\xi _{0}\ll 1$
one has

\begin{equation}
K_{s}(q)=\frac{1}{4\pi \lambda _{L}^{2}}; \chi _{n}(\mathbf{q})-\chi _{s}(%
\mathbf{q})=\chi _{n}(0)(1-\frac{\pi ^{2}q^{2}\xi _{0}^{2}}{30}),  \label{8}
\end{equation}
while for $q\xi _{0}\gg 1$ it holds

\begin{equation}
K_{s}(q)=\frac{3}{4\lambda _{L}^{2}q\xi _{0}};\chi _{n}(\mathbf{q})-\chi
_{s}(\mathbf{q})=\chi _{n}(0)\frac{\pi }{2q\xi _{0}}.  \label{9}
\end{equation}%
Based on Eqs.(12-13) and the expression for the free-energy $F$ in Eq.(8) we
obtain that just below the transition temperature $T_{m}=T_{m0}(1-3(\frac{%
\pi }{4}\frac{\theta _{ex}}{\theta }\frac{a}{\xi _{0}})^{2/3})$ a transverse
($\mathbf{Q}_{s}\mathbf{\perp }S_{z}$) \textit{sinus structure} $S_{z}(%
\mathbf{r})\approx S\sin (\mathbf{Q}_{s}\mathbf{r})$ appears. In case when $%
\xi _{0}^{2}\ll a\lambda _{L}$ the wave vector $Q_{s}$ is determined by the
EX interaction and for $\theta _{ex}/\theta _{em}\gg (a/\xi _{0})^{2}$ it is
given by \cite{BuBuKuPa}, \cite{Buzdin-Dissert}
\begin{equation}
Q_{s}=(\frac{\pi }{4}\frac{\theta _{ex}}{\theta a^{2}\xi _{0}})^{1/3}.
\label{10}
\end{equation}%
For $\theta _{ex}/\theta _{em}\ll (a/\xi _{0})^{2}$ the EM interaction
prevails
\begin{equation}
Q_{s}=(\frac{1}{a\lambda _{L}})^{1/2}.  \label{11}
\end{equation}

In the opposite limit $\xi _{0}^{2}\gg a\lambda _{L}$ the EX interaction
dominates for $\theta _{ex}/\theta _{em}\gg (a^{2}\xi _{0}/\lambda
_{L}^{3})^{2/5}$ which gives again \cite{BuBuKuPa}, \cite{Buzdin-Dissert}

\begin{equation}
Q_{s}=(\frac{\pi }{4}\frac{\theta _{ex}}{\theta a^{2}\xi _{0}})^{1/3}.
\label{12}
\end{equation}

For\ $\theta _{ex}/\theta _{em}\ll (a^{2}\xi _{0}/\lambda _{L}^{3})^{2/5}$
the EM interaction dominates which gives
\begin{equation}
Q_{s}\approx (\frac{1}{a^{2}\xi _{0}\lambda _{L}^{2}})^{1/5}\ .\   \label{13}
\end{equation}

From these expressions it is seen that in realistic cases $Q_{s}$ is
determined by the EX interaction - it does not depend on the EM parameter $%
\lambda _{L}$, while the EM interaction (with $\lambda _{L}$ dependence of
Q) is dominant only for extremely small EX interaction ($\theta _{ex}\ll
\theta _{em}$(a/$\xi _{0}$)$^{2}$ or $\theta _{ex}\ll \theta _{em}(a^{2}\xi
_{0}$/$\lambda _{L}^{3})^{2/5}$), i.e. for $(\theta _{ex}/\theta _{em})\ll
10^{-4}-10^{-5}$. However, in typical ferromagnetic superconductors, such as
ErRh$_{4}$B$_{4}$, HoMo$_{6}$S$_{8}$, HoMo$_{6}$Se$_{8}$, AuIn$_{2}$, the
\textit{EX interaction dominates} since $\theta _{ex}>0.1$ $\theta _{em}$
and $a\ll \xi _{0}\lesssim \lambda _{L}$.

In reality nonmagnetic impurities are always present and one should know $%
\chi _{s}(\mathbf{q},l)$ and $K_{s}(q,l)$ as a function of the mean-free
path $l$. \ The corresponding calculations show that if $(l^{5}/a^{2}\xi
_{0}\lambda _{L}^{2})\ll 1$ one has for $\theta _{ex}/\theta _{em}\gg
a^{2}\xi _{0}/l^{3}$ \cite{BuBuKuPa}, \cite{Buzdin-Dissert}

\begin{equation}
Q_{s}=(\frac{\pi }{4}\frac{\theta _{ex}}{\theta a^{2}\xi _{0}})^{1/3}
\label{14}
\end{equation}%
and
\begin{equation}
Q_{s}\approx \frac{\theta _{ex}}{\theta }(\frac{1}{la^{2}\xi _{0}})^{1/4}
\label{15}
\end{equation}
for\ $(a^{2}\xi _{0}/l^{3})\gg \theta _{ex}/\theta _{em}\gg (l^{2}/\lambda
_{L}^{2})$, $a^{2}l/\xi _{0}^{3}$.

In case when $(\theta _{ex}/\theta _{em})\ll (a^{2}\xi _{0}/l^{3})$ or $%
(\theta _{ex}/\theta _{em})\ll l^{2}/\lambda _{L}^{2}$ the EM interaction
dominates
\begin{equation}
Q_{s}\approx (\frac{l}{a^{2}\xi _{0}\lambda _{L}^{2}})^{1/4}.  \label{16}
\end{equation}

Let us stress some interesting properties of ferromagnetic superconductors: (%
\textbf{i}) the ferromagnetic critical temperature is strongly reduced in
the presence of SC due to the formation of Cooper pairs in the SC state,
i.e. one has $T_{F}=T_{m0}(1-\frac{\theta _{ex}+\theta _{em}}{\theta })\ll
T_{m}$. In fact this result is more general and holds also for the
coexistence of SC and \textit{itinerant ferromagnetism} (F) - singlet SC and
ferromagnetism do not coexist. In that sense a number of recent papers which
claim that the itinerant F and SC coexist should be completely abandoned
\cite{Karchev}. However, in some \textit{itinerant ferromagnets} such as $%
Y_{9}Co_{7}$ (with $T_{F}=4.5$ $K$) the microscopic parameters favor spiral
or domain magnetic structure in the SC state with $T_{c}=2.5$ $K$ as it was
proposed in \cite{BeKuSta}; (\textbf{ii}) in isotropic magnetic systems and
near the critical temperature T$_{m}$ the inverse scattering time due to
magnetic fluctuations can diverge and thus destroy SC. However, this
divergence is suppressed in the real RE ternary compounds due to the
long-range dipole-diploe interaction. The interaction of SC with magnetic
fluctuations is described by the free-energy contribution

\begin{equation}
F_{sc,fl}=\frac{\theta _{ex}}{2}\sum_{\mathbf{q}}\langle
S_{z,q}S_{z,-q}\rangle \frac{\chi _{n}(\mathbf{q})-\chi _{s}(\mathbf{q})}{%
\chi _{n}(0)},  \label{17}
\end{equation}%
where

\begin{equation}
\langle S_{z,q}S_{z,-q}\rangle \sim \frac{1}{\tau +a^{2}q^{2}+(\theta
_{em}/\theta )q_{z}^{2}/q^{2}},  \label{18}
\end{equation}%
with $\tau =(T-T_{m0})/\theta $. Due to the large dipole-dipole temperature
with $\theta _{em}\sim \theta $ these fluctuations looks like
four-dimensional, thus giving rather small value for the inverse scattering
time $\tau _{m}^{-1}\sim \theta \ll T_{c}$; (\textbf{iii}) the relative
strength of the EX and EM interaction is controlled by the parameter

\begin{equation}
r=\frac{F_{Int}^{(EM)}}{F_{Int}^{(EX)}}=\frac{\theta _{em}}{\theta _{ex}}%
\frac{1}{Q^{2}\lambda _{L}^{2}}.  \label{19}
\end{equation}
In the RE ternary compounds the case $r\ll 1$ is always realized, due to the
large value of $Q^{2}\lambda _{L}^{2}\gg 1$. Therefore, practically in all
RE ternary compounds the EX interaction dominates in the formation of
magnetic structure, while the EM interaction makes it transversal - see
exception in weak ferromagnets below; (\textbf{iv}) In spite of the fact
that in the RE ternary compounds the ferromagnetism is stronger phenomenon
than SC - the gain in the ferromagnetic energy (per LM and at T=0 K) $%
E_{m}\approx N(0)h_{0}^{2}$ is larger than the gain in the SC condensation
energy $E_{c}\approx N(0)\Delta _{0}^{2}$ since $h_{0}(\sim 10^{2}$ $K)\gg
\Delta _{0}(\lesssim 10$ $K)$, the ferromagnetic order is more "generous"
and varies spatially in the SC state. The reason lies in the fact that the
magnetic stiffness ($\sim a$) is much smaller than the superconducting
stiffness ($\sim \xi _{0}$).

\subsection{Domain magnetic structure due to SC}

By lowering $T$ the higher term $\sim S^{4}(r)$ makes the change of the
moduo of $\mathbf{S}(\mathbf{r})$ unfavorable and the sinus-structure is
transformed, as it will be shown below, into the striped \textit{domain
structure} (DS) - see $Fig.2$. At the same time since the exchange field
grows $h_{ex}=h_{0}S(T)$ but for $h_{ex}(T)<\Delta $ the mutual interaction
of magnetism and SC can be treated by the perturbation theory. In such a
case the free-energy density is completed by the density of the \textit{%
domain wall energy} $QE_{W}/\pi $, where $E_{W}$ is the wall-energy per unit
surface. In case of sufficiently large magnetic anisotropy $D_{z}>\theta $
rotation of the moments in the wall is unfavorable and the \textit{linear
domain wall} is favored, i.e. $S_{z}(x)=Sth(x/l_{W})$, $S_{x}=S_{y}=0$,
where $l_{W}=a/\sqrt{\tau }$ is the domain-wall thickness \cite%
{BuBuKuPaAdvances}. The domain wall energy per unit surface is given by
\begin{figure}[tbp]
\resizebox{.6 \textwidth}{!} {
\includegraphics*[width=8cm]{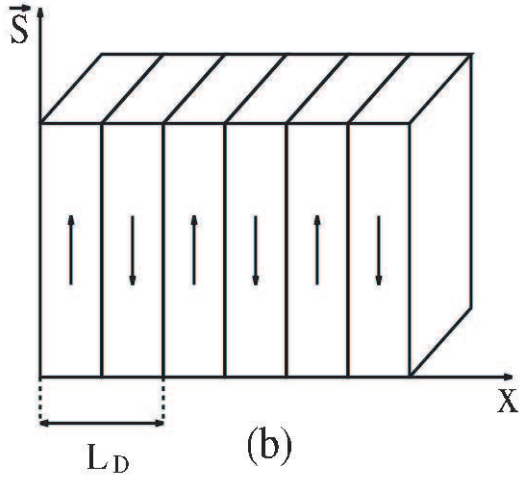}}
\caption{The striped domain magnetic structure $\mathbf{S}(x)=S_{z}(x)%
\mathbf{e}_{z}$ with the period $L_{D}=2\protect\pi /Q_{DS}$; $\mathbf{Q}$
is along the x-axis.}
\end{figure}
\begin{equation}
E_{W}=(4\sqrt{2}/3)n\theta S^{2}a\tau ^{1/2}\equiv n\theta S^{2}\tilde{a}
\label{20}
\end{equation}%
where $\tau =(T-T_{m0})/\theta $) - see \cite{BuBuKuPaAdvances}.The
free-energy density $\tilde{F}_{DS}$ in the DS phase is
\begin{equation}
\tilde{F}_{DS}=n\theta \lbrack \frac{1}{2}\tau S^{2}+\frac{b}{4}S^{4}]+\frac{%
Q}{\pi }E_{W}+n\theta _{ex}\frac{7\zeta (3)}{2\pi }\frac{S^{2}}{Q\xi _{0}}.
\label{21}
\end{equation}%
Note, that if the anisotropy energy is small, i.e. $\tau >2D_{z}/\theta $ \
the \textit{rotating domain wall} is realized with the wall thickness $%
l_{W}\approx a(\theta /D_{z})^{1/2}$ and the wall energy $E_{W}\sim
S^{2}a(\theta D_{z})^{1/2}$. In the case of an extremely small anisotropy$%
(D_{z}/\theta )<(a/\xi _{0})\sim 10^{-2}-10^{-3}$ then the spiral structure
is realized. Minimizing $F_{DS}$ in Eq.(25) with respect to $Q$ one obtains
the equilibrium wave vector of the striped DS phase
\begin{equation}
Q_{DS}\simeq 2(\frac{\theta _{ex}}{\theta }\frac{1}{\tilde{a}\xi _{0}})^{1/2}
\label{22}
\end{equation}%
We conclude that the period of the DS structure is larger than for the sinus
phase. It is worth of mentioning that: (i) the here obtained striped domain
structure is due to SC and it is property of the bulk, (ii) the problem of
the DS phase in SC is mathematically similar to the problem of domain
structure in a normal ferromagnetic plate with the magnetization
perpendicular to the plat plane, where the role of $F_{int}$ is played by
the magnetic energy dissipated out of plate. Generally, the domain structure
is realized when the wall thickness $a/\sqrt{\tau }$ is much smaller then
the striped domain thickness $\pi /Q$ implying that $\tau \gg (a/\xi
_{0})^{2/3}\sim 10^{-2}$.

At lower temperatures when $h_{ex}(T)>\Delta $ the nonperturbative problem
is studied in the presence of impurities by the \textit{quasiclassical ELO}
equations (see Appendix 7.2) since the period $L_{D}$ of the domain
structure is much larger than $a$, i.e. $L_{D}\gg a$. This equations are
solved for the domain structure with%
\begin{equation}
S_{z}(\mathbf{r})=\frac{4S(T)}{\pi }\sum_{k=0}^{\infty }\frac{\sin (2k+1)%
\mathbf{Qr}}{2k+1}\equiv \sum_{\mathbf{q}}S_{z,\mathbf{q}}e^{i\mathbf{qr}}.
\label{23}
\end{equation}%
By assuming that $\mathbf{Q}$ is along the x-axis the solution for the
Green's function are searched in the form

\begin{equation}
f(\mathbf{v},x)=f_{0}(\mathbf{v})+\sum_{k}f_{k}(\mathbf{v})e^{ikx},
\label{24}
\end{equation}%
and analogously for $g(\mathbf{v},x)$, where $k=(2m+1)Q$ and $\mid f_{k}\mid
\ll \mid f_{0}\mid $, $\mid g_{k}\mid \ll \mid g_{0}\mid $. The calculations
were done in \cite{BuBuKuPa}, \cite{BuBuKuPaAdvances} and here we present
only the final result for the free-energy in the \textit{dirty limit} ($l\ll
\xi _{0}$). It turns out, that in that case the interaction of the magnetic
domain structure with SC is similar to the case of magnetic impurities with
the inverse scattering time $\tau _{m}^{-1}$ and with $\tau _{m}\Delta >1$,
i.e. $\tilde{F}_{DS}$ is given by
\[
\tilde{F}_{DS}=n\theta \lbrack \frac{1}{2}\tau S^{2}+\frac{b}{4}%
S^{4}]+QE_{W}
\]%
\begin{equation}
-\frac{1}{2}N(0)\Delta ^{2}\ln (\frac{e\Delta _{0}^{2}}{\Delta ^{2}})+N(0)%
\frac{\pi \Delta }{2\tau _{m}}(1-\frac{2}{3\pi \tau _{m}\Delta }).
\label{25}
\end{equation}%
$\tau _{m}^{-1}$ is given by ($h_{z,q}=h_{0}S_{z,\mathbf{q}}$)

\begin{equation}
\tau _{m}^{-1}=\sum_{q}\{\frac{\pi h_{z,q}h_{z,-q}}{v_{F}q}L_{1}(ql)+\frac{3%
\mathbf{B}_{q}\cdot \mathbf{B}_{-q}}{16\lambda _{L}^{2}nN(0)v_{F}q^{3}}%
L_{2}(ql)\}  \label{26}
\end{equation}%
where

\begin{equation}
L_{1}(y)=\frac{2y\arctan y}{\pi (y-\arctan y)}
\end{equation}%
and

\begin{equation}
L_{2}(y)=\frac{2}{\pi }[(1+\frac{1}{y^{2}})\arctan y-\frac{1}{y}].
\label{27b}
\end{equation}
The magnetic induction $\mathbf{B}_{q}$ is given by

\begin{equation}
\mathbf{B}_{q}=\frac{4\pi n\mu \lbrack q^{2}\mathbf{S}_{q}-\mathbf{q}(%
\mathbf{qS}_{q})]}{q^{2}+K_{s}(q)(1-4/3\pi \tau _{m}\Delta )},  \label{27}
\end{equation}%
where the Kernel $K_{s}(q)$ in the dirty limit has the form

\begin{equation}
K_{s}(q)=\frac{3\pi \Delta }{16v_{F}\lambda _{L}^{2}q}L_{2}(y).  \label{27b}
\end{equation}

Based on the free-energy in Eq.(29) we can study the coexistence problem in
the whole temperature regions and for various $Ql$ - see more in \cite%
{BuBuKuPa}. We summarize the main results: \textit{(i)} at $T=T_{m}$ the
sinusoidal magnetic order appears with the wave vector $Q_{s}\sim
(1/a^{2}\xi _{0})^{1/3}$ ; \textit{(ii)} by lowering temperature the domain
structure appears with $Q_{DS}\sim (1/a\xi _{0})^{1/2}$ which persists up to
the temperature of the \textit{first order phase transition} $T_{c2}$ where
the DS phase passes into the normal ferromagnetic state. At $T_{c2}$ one has

\begin{equation}
F_{DS}\{S_{DS}(T_{c2}),\Delta
(T_{c2}),Q_{DS,c2}\}=F_{FN}\{S_{F}(T_{c2}),0,0\},  \label{28}
\end{equation}%
where $Q_{DS,c2}\approx 1.8(\tilde{a}(T_{c2})\xi _{0})^{-1/2}\sim (a\xi
_{0})^{-1/2}$, $\Delta (T_{c2})=0.85\Delta _{0}$ and $(S_{c2}^{2}/Q_{c2})$ $%
\approx 0.07(\Delta _{0}v_{F}/h_{0}^{2})$; \textit{(iii)} if $%
S_{DS}(T_{c2})>1$ then the DS is stable up to $T=0$ $K$ - this situation is
realized in systems with small EX interaction (which still dominates over
EM), i.e. for $\theta _{ex}<$ $\theta _{ex}^{c}\sim (T_{c}^{3}/h_{0}^{2})$; (%
\textit{iv}) in \textit{dirty} SC with $(h\tau )^{2}\ll 1$ there is a gap in
the quasiparticle spectrum for $E<\Delta $ in the whole range of the
existence of the domain phase. The calculations in \textit{clean} SC show
\cite{BuKuRu}, \cite{BuBuKuPaAdvances} that when $h(T)\gg \Delta $ the
\textit{spectrum is gapless}. For instance in the \textit{DS phase} one has
for $E\ll \Delta $

\begin{equation}
\frac{N(E)}{N(0)}=\frac{\pi h}{v_{F}Q}\frac{E}{\Delta }\ln \frac{4\Delta }{E}%
,  \label{28a}
\end{equation}%
while in the case of the \textit{spiral order}

\begin{equation}
\frac{N(E)}{N(0)}=\frac{\pi h}{2v_{F}Q}\frac{E}{\Delta }.  \label{28b}
\end{equation}%
(\textit{v}) The spin-orbit interaction decreases the value of $\chi _{n}(%
\mathbf{0})-\chi _{s}(\mathbf{0})$ (and small $q$) and it is detrimental for
the DS phase. However the analysis in \cite{SO-BuBuKu} shows that the S-O
scattering destroys the peak in $\chi _{s}(\mathbf{q})$ only in very dirty
systems when $l\sim a$.

We would like to point out, that there were a lot of studies of
ferromagnetic superconductors based on the phenomenological theory which
takes into account the EM interaction only \cite{Krey}, \cite{Blount}.
However, this interesting phenomenology is inadequate in describing real
materials, such as the above numbered RE ternary compounds where the EX
interaction prevails in the formation of the oscillatory structure (with Q$%
\gg \xi _{0}^{-1}$, $\lambda _{L}^{-1}$ ) in the SC state. We stress that in
the above theory the EM interaction, as well as the EX one, is taken into
account on the microscopic level, thus giving much more reliable predictions
than the phenomenological approach.

\subsubsection{Experimental situation}

In the most important ferromagnetic superconductors HoMo$_{6}$S$_{8}$, ErRh$%
_{4}$B$_{4}$, HoMo$_{6}$Se$_{8}$ the range of microscopic parameters allows
the coexistence of SC and modified ferromagnetic order. For instance in
clean systems one has \cite{BuBuKuPaAdvances}: in\textbf{\ ErRh}$_{4}$%
\textbf{B}$_{4}\ $- $\ n\sim 10^{22}$ $cm^{-3}$, $\mu =5.6$ $\mu _{B}$, $%
\tilde{a}\approx 1$ $\mathring{A}$, $\lambda _{L}(0)\approx 900$ $\mathring{A%
}$, $\xi _{0}\approx 200$ $\mathring{A}$, $\Delta _{0}\approx 15.5$ $K$, $%
N^{-1}(0)=1850$ $K\cdot spin$ $RE$, $v_{F}\approx 1.3\times 10^{7}$ $%
cm^{-1}s $, $\theta _{ex}\approx 0.5$ $K$, $h_{0}\approx 40$ $K$, $\tau
_{m}^{-1}\approx 3$ $K$ and $\theta _{em}\approx 1.8$ $K$; in \textbf{HoMo}$%
_{6}$\textbf{S}$_{8}$ - $n\sim 4\times 10^{21}$ $cm^{-3}$, $\mu =9.1$ $\mu
_{B}$, $\tilde{a}\approx 2.5$ $\mathring{A}$, $\lambda _{L}(0)\approx 1200$ $%
\mathring{A}$, $\xi _{0}\approx 1500$ $\mathring{A}$, $\Delta _{0}\approx
3.2 $ $K$, $N^{-1}(0)=3600$ $K\cdot spin$ $RE$, $v_{F}\approx 1.8\times
10^{7}$ $cm^{-1}s$, $\theta _{ex}\approx 0.2$ $K$, $h_{0}\approx 24$ $K$, $%
\tau _{m}^{-1}\approx 0.9$ $K$ and $\theta _{em}\approx 1.3$ $K$ while in
HoMo$_{6}$Se$_{8}$ a number of parameters are similar to HoMo$_{6}$S$_{8}$.

An oscillatory magnetic structure (either sinus or domain-like) due to SC
has been observed at least in three compounds: \textbf{(1)} in HoMo$_{6}$S$%
_{8}$ where $T_{c}=8.7$ $K$, $T_{m}\approx 0.8$ $K$, $T_{c2}\approx 0.7$ $K$
and $Q_{DS}\sim 0.03$ $\mathring{A}^{-1}$; \textbf{(2)} in ErRh$_{4}$B$_{4}$
where $T_{c}=8.7$ $K$, $T_{m}\approx 0.8$ $K$, $T_{c2}\approx 0.7$ $K$ and $%
Q_{DS}\sim 0.06$ $\mathring{A}^{-1}$; \textbf{(3) }in HoMo$_{6}$Se$_{8}$
where $T_{c}=5.5$ $K$, $T_{m}=0.5$ $K$ and $Q_{DS}\sim 0.09-0.06$ $\mathring{%
A}^{-1}$ and the coexistence persists up to $T=0$ $K$! All these results are
in a satisfactory agreement with the above theory. The conclusion is that in
most RE ternary compounds the EX interaction is responsible for the
formation of the oscillatory magnetic structure in the SC state, while the
EM interaction makes the structure transverse $\mathbf{Q\cdot S}(\mathbf{r}%
)=0$.

\subsection{Domain magnetic structure in thin SC film}

In the above calculations we have assumed that the thickness $L$ of the
sample is very large, i.e. $L\gg \xi _{0}$, so that the dissipated magnetic
energy (stray field) can be neglected. In case of thin films with $L\sim \xi
_{0}$ the stray\ magnetic energy $E_{st}$ existing around the domain walls
must be added to the free-energy $F_{DS}$ in Eq.(29) (or its simple version
in Eq.(25)), i.e. $F_{tot}=F_{DS}+E_{st}$ is given by \cite{Buzdin-Dissert}

\begin{equation}
\tilde{F}_{tot}/n=(\tilde{F}_{DS}/n)+E_{st}=(\tilde{F}_{DS}/n)+0.85\theta
_{em}\frac{S^{2}(T)}{QL}.  \label{28c}
\end{equation}%
In case when the ratio $r(=F_{Int}^{(EM)}/F_{Int}^{(EX)})\ll 1$ the
minimization of $F_{tot}$ w.r.t. $Q$ gives

\begin{equation}
Q_{tot}^{2}=Q_{DS}^{2}+Q_{F}^{2},  \label{28d}
\end{equation}%
where $Q_{DS}$ is the wave vector of the DS phase without stray magnetic
energy and $Q_{F}\approx 1.6(\theta _{em}/\theta \tilde{a}L)^{1/2}$ is the
wave vector of the domain structure in the normal ferromagnetic state. From
Eq.(39) it is seen that the period of the DS ($d=2\pi /Q_{tot}$) in thin SC
film is decreased due to the stray field. It comes out from Eq.(38) that the
transition temperature $T_{c2}$ (for the first order phase transition $%
DS\rightarrow FN(domain)$ can be pushed to zero when $L<L_{c}=3\xi
_{0}(\theta _{em}S_{c2}^{4}(L=\infty ))/\theta _{ex}(1-S_{c2}^{4}(L=\infty
))^{2}$. \ The experiments on thin films of HoMo$_{6}$S$_{8}$ show such a
thickness dependence of $T_{c2}$ where $T_{c2}(L)<T_{c2}(\infty )$.

Let us mention that even in the normal ferromagnetic state, which is
realized for $T<T_{c2}$, there is possibility that SC exist in the domain
walls as it was shown in \cite{Kopaev}, \cite{Kulic-Domain}, \cite%
{BuBu-Domain}, \cite{BuBuKuPaAdvances}. This situation can be realized in
some pseudoternary compounds where $h_{0}\lesssim \Delta _{0}$.

\subsection{Coexistence of Nuclear Magnetism and Superconductivity}

In 1997 the Pobell's group from Bayreuth made an important discovery \cite%
{Pobell} by observing that \textit{superconductivity and nuclear magnetism
coexist} in AuIn$_{2}$ with $T_{c}=0,207$ $K$ and $T_{m}=35$ $\mu K$. At
first glance this is not too surprising having in mind smallness of the
hyperfine interaction between conduction electrons and nuclear spins.
However Buzdin, Bulaevskii and the present author applied in 1997 \cite%
{KuBuBu} the theory of magnetic superconductors \cite{BuBuKuPaAdvances} and
found a surprising result - the effective nuclear "exchange" field (the
hyperfine contact interaction) is rather large $h_{hyp}$ $\approx 1$ $K$
while $\Delta _{0}\approx 0.6$ $K$, i.e. $h_{hyp}>\Delta _{0}$! The \textit{%
hyperfine interaction} has the same (mathematical) structure as the exchange
interaction between the 4f LMs and conduction electrons
\begin{equation}
\hat{H}_{e-nuc}=\int d^{3}r\sum_{i}A_{hyp}\delta (\mathbf{r}-\mathbf{R}_{i})%
\hat{\psi}^{\dagger }(\mathbf{r})\mathbf{\sigma \hat{I}}_{i}\hat{\psi}(%
\mathbf{r})  \label{29}
\end{equation}%
Here, $A_{hyp}$ is the hyperfine interaction and the "exchange field" is
given by $h_{hyp}=nA_{hyp}\langle $ $\mathbf{\hat{I}}_{i}\rangle $, where $%
\mathbf{\hat{I}}_{i}$ is the nuclear spin. So, the nuclear magnetism in $%
AuIn_{2}$, which shows strong tendency toward ferromagnetism, competes
rather strongly with SC. It was estimated from the experiment \cite{Pobell}
that $\theta _{em}(=2\pi n_{n}\mu _{n}^{2})\approx 1$ $\mu K$ and $\theta
_{ex}(\approx N(0)h_{hyp}^{2})\approx 35$ $\mu K$, $\xi _{0}\approx 10^{5}$ $%
\mathring{A}$, $\lambda _{L}\approx 10^{5}$ $\mathring{A}$, $l\approx
3.6\times 10^{4}$ $\mathring{A}$ $(l<\xi _{0}$) which means that the "EX"
(contact) interaction is much stronger than the EM (dipole-dipole) one and
the theory invented for RE ternary compounds is completely applicable to
this case. This theory predict, that if the nuclear magnetic anisotropy (due
to the dipole-dipole interaction) is small, i.e. $(D/\theta _{ex})<10^{-3}$,
the spiral magnetic structure should be realized, while in the opposite case
$(D/\theta _{ex})>10^{-3}$ the striped domain structure is formed. The
experiments in magnetic field \cite{Pobell} give evidence that SC and
oscillating magnetic order coexist up to $T=0$ $K$, i.e. the case $\theta
_{ex}<\theta _{ex}^{c}$ is realized in the type-I superconductor $AuIn_{2}$.
Unfortunately, until now there were no nuclear scattering measurements on $%
AuIn_{2}$ which could resolve the nuclear magnetic structure below $T_{m}=35$
$\mu K$.

The study of the coexistence of SC and nuclear magnetic order is of enormous
importance for the fundamental physics. These systems give an opportunity to
study the coexistence problem in cases when the electronic temperature ($%
T_{e}$) is different than the nuclear one ($T_{n}$), i.e. $T_{e}\neq T_{n}$.
But probably the most interesting problem is the coexistence of SC and
nuclear magnetism in the case of \textit{negative nuclear temperatures }($%
T_{n}<0$ $K$).

\section{Antiferromagnetic superconductors (AFS)}

\subsection{Coexistence of antiferromagnetism and superconductivity}

An evident experimental fact in the RE ternary compounds is that
superconductivity coexists with the antiferromagnetic (AF) order much easier
than with the modified ferromagnetic order. The reason is that the effective
exchange field in AFS varies on the lattice constant (the AF\ wave vector is
$Q_{AF}\sim a^{-1}$) and it is averaged to zero over the volume of the
Cooper pair $\xi _{0}^{3}$. Thinking in terms of the electronic
susceptibility one has

\begin{equation}
\frac{\chi _{n}(\mathbf{Q}_{AF})-\chi _{s}(\mathbf{Q}_{AF})}{\chi _{n}(0)}%
\approx \frac{\Delta }{v_{F}Q_{AF}}\sim \frac{T_{c}}{E_{F}}\ll 1,  \label{30}
\end{equation}

which means that $F_{Int}^{(EX)}$ in AFS\ is very small. Due to the same
reason the EM interaction is small since $\delta K_{s}(Q_{AF})\sim
a^{3}/(\lambda _{L}^{2}\xi _{0})$, i.e. $F_{Int}^{(EM)}(\ll F_{Int}^{(EX)})$
This result is confirmed in a number of RE ternary compounds in which the
Neel temperature $T_{N}(\approx N(0)h^{2})$ is in most cases (much) smaller
than $T_{c}$ \cite{Maple}. In these systems the magnetic scattering above $%
T_{N}$ is not harmful for SC since the inverse life time $\tau _{m}^{-1}\sim
T_{N}\ll T_{c}$.

However, there are a number of interesting properties of the AF
superconductors (AFS) such as the pair-breaking effect of nonmagnetic
impurities characterized by the life-time $\tau $. In case when $T_{N}\ll
T_{c}$ the nonmagnetic impurities in the AFS state are pair-breaking, like
magnetic impurities with the inverse scattering time

\begin{equation}
\tau _{m}^{-1}=\frac{\pi h^{2}}{2v_{F}Q_{AF}\sqrt{1+(h\tau )^{2}}}.
\label{31}
\end{equation}%
For $h\tau \ll 1$ one gets $\tau _{m}^{-1}\sim T_{N}$ $\ll T_{c}$ which
means that in this case the pair-breaking effect of impurities is rather
small \cite{Morozov}. Very interesting situation appears for systems with $%
T_{N}$ $\gg T_{c}$. Even in such a case the exchange field does not suppress
$T_{c}$ significantly since the theory (based on Eqs.(63-64) in Appendix $%
7.1 $) predicts that $(\delta T_{c}/T_{c0})\sim (h/E_{F})(\ln h/E_{F})\ll 1$%
. However, in the presence of nonmagnetic impurities $T_{c}$ is renormalized
appreciably and SC disappears \ for $l<l_{c}\approx 10\xi
_{0}(h/v_{F}Q_{AF})\sim \xi _{0}T_{N}/h$. In that respect there is one very
interesting AFS compound $Tb_{2}Mo_{3}S_{4}$ with $T_{N}=19$ $K$ and $%
T_{c}=0.8$ $K$ where one expects that SC should disappear due to the strong
magnetic scattering. However, it turns out that in this system the magnetic
anisotropy, in conjunction with large momentum $J=9$, strongly suppress this
pair-breaking effect giving rise for SC.

\subsection{Weak ferromagnetism in AFS}

In the case of competition of SC and the ferromagnetic order in the RE
ternary compounds the theory predicts that in the presence of an appreciable
EX interaction SC\ can coexist only with spiral and DS (or sinus) order -
depending on the magnetic anisotropy, while other phases are excluded. It
turns out that in \textit{AF superconductors with weak ferromagnetism (WF)}
- of the Moriya-Dyalozhinski type, the phase diagram can be much reacher
containing also the \textit{Meissner phase} ($M\neq 0$, $B=0$) and the
\textit{spontaneous vortex state} \cite{Buzdin-Krotov}. We discuss this
problem briefly by studying the AF order with two sublattice where the AF
order parameter is given by $\mathbf{l}=\mathbf{S}_{1}-\mathbf{S}_{2}$. In
systems which allow WF there is an additional term in the free-energy $%
F_{WF}=\mathbf{D[\mathbf{S}_{1}\times \mathbf{S}_{2}]}$ in the total
free-energy. If for instance $\mathbf{l}$ is along the xy-plane and $\mathbf{%
D}$ is so oriented that it allows the appearance of the weak ferromagnetism $%
\mathbf{m}=\mathbf{S}_{1}+\mathbf{S}_{2}$ ($\mathbf{M}=n\mu \mathbf{m}$) in
the xy-plane then $F_{WF}$ is given by

\begin{equation}
\tilde{F}_{WF}=\beta n\theta _{ex}(m_{x}l_{y}+m_{y}l_{x}).  \label{32}
\end{equation}%
Since in most systems $m\sim 10^{-3}l$ it immediately implies that $\beta
\ll 1$. In that case and when $T_{N}$ $\ll T_{c}$ the interaction part $%
F_{int}$ of the total free-energy ($F=F_{m}+F_{s}+F_{int}$) is given by
Eq.(7), while the magnetic system is described by $F_{m}$
\[
F_{m}=\int d^{3}rn\theta _{ex}[a_{l}\mathbf{l}^{2}+\frac{c}{4}(\mathbf{l}%
^{2})^{2}+b\mathbf{m}^{2}+a^{2}(\nabla \mathbf{l)}^{2}]\
\]

\begin{equation}
+\int d^{3}r[\beta n\theta _{ex}(m_{x}l_{y}+m_{y}l_{x})+\frac{(\mathbf{B}%
-4\pi \mathbf{M})^{2}}{8\pi }].  \label{33}
\end{equation}

By minimizing $F$ w.r.t. $\mathbf{A},\mathbf{l},m$ and $\mathbf{q}$ we get
possible phases in AFS with WF \cite{Buzdin-Krotov}. The resulting
free-energy is similar to the case of ferromagnetic superconductors with an
effective magnetic stiffness $a_{eff}=(ab/\beta )\gg a$. It turns out that
if $\beta \gg a/\xi _{0}$ the EX interaction dominates in the formation of
the magnetic structure, and the sinus structure ($l\sim \sin \mathbf{Qr} $
and $m\sim \sin \mathbf{Qr}$) appears at T$_{N}$, while for $(a/\lambda
_{L})<\beta \ll a/\xi _{0}$ the EM interaction prevails in the formation of
the sinusoidal structure. If $\beta <(a/\lambda _{L})\sqrt{2\theta
_{em}/\theta _{ex}}$ than the nonuniform structure is unfavorable and the so
called \textit{Meissner state} (first proposed by Ginzburg in 1956) appears.
It is characterized by $\mathbf{M=}const$ and $\mathbf{B}=0$ in the bulk
sample due to the screening SC current on the surface ($B=4\pi M\exp
\{-z/\lambda _{L}\}$). By lowering the temperature $\mid \mathbf{S}_{1,2}$ $%
\mid $ grow and it is necessary to take into account higher order terms in $%
F $. As a result one gets that for $\beta \gg \sqrt{a/\xi _{0}}$ again the
EX dominates and the striped DS phase is realized, while for $\sqrt{%
a/\lambda _{L}}\ll \beta \ll \sqrt{a/\xi _{0}}$ the striped DS phase is
realized due to the EM interaction. However, by lowering the temperature the
domain wall energy grows and it my happen that a \textit{spontaneous vortex
state} (with $4\pi M>H_{c1}$ - the lower critical field) appears for $\beta
\ll \sqrt{a/\xi _{0}}$ and for the AF\ vector $l>l_{c}\sim
(H_{c1}/M(0))(\beta \lambda _{L}^{2}/\tilde{a}^{2})^{1/3}$, $\tilde{a}%
=a[(T_{N}-T)/T_{N}]^{1/2}$. From the known RE ternary compounds the good
candidate for such a behavior is the body centered tetragonal (b.c.t.)
system $ErRh_{4}B_{4}$.

\section{Magnetic superconductors in magnetic field}

There are a number of interesting effects of the magnetic field $H$ either
in the coexistence phase or above the magnetic transition temperature $T_{m}$
where $S(T)=0$. \ We discuss some of them briefly - for more details see
\cite{BuBuKuPaAdvances}, \cite{SO-BuBuKu}.

(1) \textit{DS phase in magnetic field} - In the case of bulk samples
magnetic field penetrates only on the length $\lambda _{L}$, thus affecting
surface of the sample only. However, in thin films the paramagnetic effect
of the field is most important \cite{BuBuKuPaAdvances}. This problem was
studied in the case of a thin (along the y-axis) film the thickness $%
L_{y}<\xi _{0}$ when the magnetic field is parallel to the domains, i.e. $%
\mathbf{H}=H\mathbf{e}_{z}$ - see Fig. 2. It is found that the magnetization
$S_{z}(x)$ contains besides the odd harmonics also the zeroth-one as well as
the even ones

\[
S_{z}(x)=S\delta +\sum_{k=1}^{\infty }\frac{2S}{\pi k}\{[1-(-1)^{k}\cos (\pi
k\delta )]\sin (kQx)
\]

\begin{equation}
+(-1)^{k}\sin (\pi k\delta )\cos (kQx)\},  \label{33a}
\end{equation}%
with $\delta =\mu H/2S\theta _{ex}$. This change of harmonics in $S_{z}(x)$
can be observed by magnetic neutron diffraction experiments. The result in
Eq. (45) means that the domains with $\mathbf{M}$ parallel to $\mathbf{H}$
increase their thickness $d\rightarrow d(1+\delta )$ while those
antiparallel decrease it $d\rightarrow d(1-\delta )$. In case when the
zeroth component of the exchange field $\bar{h}(=h_{0}S\delta )>\bar{h}%
_{c}=\Delta \lbrack 1-(1/\tau _{m}\Delta )^{2/3}]^{2/3}$ the DS phase is
destroyed due to the Zeeman effect making $\Delta =0$. For $\bar{h}<\bar{h}%
_{c}$ the parameters of the DS phase are renormalized, i.e. $Q(H)<Q(0)$. In
case when $\mathbf{H}=H\mathbf{e}_{y}$ (i.e. orthogonal to the z-axis) then
all domains have the same thickness and there is no redistribution of
intensities of neutron peaks. However, there is only a decrease of
intensities of (2k+1)Q peaks by the factor ($1-\delta _{\perp }^{2}$) where $%
\delta _{\perp }=\mu H/S(\theta _{ex}+D_{z})$ \ and $D_{z}$ is the magnetic
anisotropy.

(2) \textit{MS in magnetic field at} $T>T_{m}$ - The effect of the exchange
field on SC in magnetic field is negligible for $T\lesssim T_{c}$ since for $%
T_{m}\ll T_{c}$ the magnetic susceptibility $\chi _{m}$ is very small.
However, at $T$ near $T_{m}$ there is a significant increase of $\chi _{m}$
and accordingly the increase of the paramagnetic effect.

(\textbf{i}) \textit{Thermodynamic critical field} $H_{c}(T)$ - We
illustrate this effect by analyzing the change of the thermodynamical field $%
H_{c}(T)$ (for the transition $N\rightarrow MS$) in magnetic
superconductors. In that case the Gibbs energy density of the paramagnetic
normal phase is equal to that of the SC phase, $\tilde{G}_{N}(H_{c})=\tilde{G%
}_{SC}(H_{c})$ where

\begin{equation}
\tilde{G}_{SC}(H_{c})=\tilde{F}_{n}(0)-\frac{H_{c0}^{2}}{8\pi }  \label{33b}
\end{equation}

\begin{equation}
\tilde{G}_{N}(H_{c})=F_{n}(0)-\frac{\mu H_{c}^{2}}{8\pi }  \label{33c}
\end{equation}%
which gives the critical field

\begin{equation}
H_{c}(T)=\frac{H_{c0}(T)}{\sqrt{1+4\pi \chi _{m}(T)}}.  \label{33d}
\end{equation}

Here, $(H_{c0}^{2}/8\pi )=N(0)\Delta ^{2}/2$ is the SC condensation energy
and the magnetic permeability is $\mu =1+4\pi \chi _{m}$ (here we neglect
the conduction electron susceptibility $\chi _{e}$ since in MS one has $\chi
_{e}\ll \chi _{m}$). For $T>T_{m}$ one has $\chi _{m}(T)\approx (\theta
_{em} T_{m0}/4\pi \theta )/(T-T_{m0})$ and
\begin{equation}
H_{c}(T)\sim \sqrt{T-T_{m0}}  \label{33e}
\end{equation}%
i.e. $H_{c}(T)$ is drastically reduced near $T_{m0}$.

(\textbf{ii}) \textit{Upper critical field} $H_{c2}(T)$ - Above $T_{m0}$ in
the presence of the \textit{external field} $\mathbf{H}_{e}$
superconductivity is suppressed by the orbital effect of the field $\mathbf{B%
}=\mathbf{H}_{i}(1+4\pi \chi _{m})$ and by the paramagnetic effect of the
exchange field $\mathbf{h}$ (and by the much smaller effect due to $\mathbf{B%
}$). Here, $\mathbf{H}_{i}=\mathbf{H}_{e}+\mathbf{H}_{D}$ where $\mathbf{H}%
_{D}$ is the demagnetization field. The critical field can be calculated by
the same formula as for usual SC - see \cite{WHH-Hc2}, where $\mu _{B}$ is
replaced by $\tilde{\mu}_{B}=\mu _{B}+h_{0}M/n\mu H_{i}$ and the electron
charge $e$ by $\tilde{e}=e(1+4\pi M/H_{i})$. In the pure limit and for $T\ll
T_{c}$ one gets the modified Gruenberg-G\"{u}nther formula \cite{Gru-Gu-para}

\begin{equation}
H_{c2}(T)=\frac{\sqrt{2}}{1+4\pi \chi _{m}(T)}H_{c2}^{\ast }(0)\frac{%
f(\alpha )}{\alpha }  \label{33f}
\end{equation}%
where $H_{c2}^{\ast }(0)$ us the upper orbital critical field in absence of
magnetic moments and $f(\alpha )$ is calculated numerically in \cite%
{Gru-Gu-para}. The parameter $\alpha$ describes the relative role of the
orbital and paramagnetic effect

\begin{equation}
\alpha =\frac{2H_{c2}^{\ast }(0)h_{0}\chi _{m}(T)}{(1+4\pi \chi _{m}(T))n\mu
\Delta _{0}}.  \label{33g}
\end{equation}

In the RE ternary magnetic superconductors one has $h_{0}\gg \Delta _{0}$
and $n\mu $ is one order of magnitude smaller than $H_{c2}^{\ast }(0)$ thus
giving $\alpha \gg 1$ in the region where $T\ll T_{c}$. It is known that for
$\alpha >1.8$ \cite{Gru-Gu-para} in pure superconductors it is realized the
Larkin-Ovchinikov-Fulde-Ferrell (LOFF) phase (due to paramagnetic effects)
where the SC\ order parameter oscillates being also zero at some points. For
$\alpha \gg 1$ one has $f(\alpha )\approx 1$ and

\begin{equation}
H_{c2}(T)\approx 1.5\frac{\Delta _{0}T_{m0}}{h_{0}\mu \theta }(T-T_{m0})
\label{33h}
\end{equation}%
i.e. $H_{c2}(T)$ depends linearly on $T-T_{m0}$ near $T_{m0}$, which is much
faster falloff than of $H_{c}(T)$. This leads to an very interesting effect
that by the first order phase transition at $H_{c}(T)$ the system goes into
the Meissner or vortex state depending on the relation between $H_{c}$ and $%
H_{c1}$. Let us mention that the lower critical field $H_{c1}$ is very
weakly affected by the exchange field due to the localized moments. The
theory \cite{BuBuKuPaAdvances} predicts the following dependence of $H_{c1}$

\begin{equation}
H_{c1}=\frac{\Phi _{0}}{4\pi \lambda _{L}^{2}}\ln \frac{\lambda _{L}\sqrt{p}%
}{\xi },  \label{low}
\end{equation}
where $p$ takes into account screening effects due to EX and EM interaction

\begin{equation}
p(T)=1-\frac{\theta _{em}}{\theta _{em}+\theta _{ex}+\frac{\theta (T-T_{m0})%
}{T_{m0}}}.  \label{p}
\end{equation}%
Note, the theory based on the EM interaction (which assumes $\theta _{ex}=0$%
) gives $p(T\rightarrow T_{m0})=0$, which makes the effective penetration
depth zero, i.e. $\lambda _{eff}=\lambda _{L}\sqrt{p}=0$, and the
Ginzburg-Landau parameter $\kappa =(\lambda _{eff}/\xi )\rightarrow 0$. If
this would be correct than we would have change from type-II SC to type-I SC
near $T_{m0}$. This result is apparently incorrect in the RE ternary
compounds, since $\theta _{ex}\sim \theta _{em}$, thus making $p$ finite and
$\kappa $ stays practically unchanged. So, the change of the type of
transition near $T_{m0}$ is not due to the change of $\kappa $ but it is due
to the much faster temperature falloff of $H_{c2}(T)$ than of $H_{c}(T)$. We
shall not discuss further this interesting subject but refer the reader to
\cite{BuBuKuPaAdvances} where various phases in the H-T phase diagram were
analyzed. Depending on the demagnetization effects several phases can be
realized in the same material, such as Meissner-, vortex-, LOFF- or even
intermediate-phase.

\section{Josephson effect on magnetic superconductors}

After the remarkable theoretical discovery by Buzdin and coworkers of the
possibility of $\pi $-Josephson junctions in the hybrid $S-F-S$ structure
where F is a ferromagnet \cite{Buzdin-pi-shift}, \cite{Buzdin-Review} the
interest in Josephson junctions with magnetic degrees of freedom has grown
dramatically - see this issue. In that sense it is natural challenge to
investigate this problem in magnetic superconductors.

\subsection{$\protect\pi $-contact due to triplet amplitude $F_{\uparrow
\uparrow }$ $(F_{\downarrow \downarrow })$}

In \cite{Kic-Ig} the tunnelling Josephson junction was studied with the left-%
$L$ and right-$R$ bulk magnetic superconductors in which the \textit{spiral
magnetic ordering }is realized - see $Fig.3$. The spiral magnetic order is
characterized by the wave vector $\mathbf{Q}_{L,R}$ and the exchange fields $%
h_{\theta _{L(R)}}=h_{L(R)}e^{i\theta _{L(R})}$, respectively, while
superconductivity in the banks\ is described by the order parameter $\Delta
_{L,R}=\mid \Delta _{L,R}\mid e^{i\varphi _{L,R}}$. Due to simplicity it was
assumed $\mid \Delta _{L}\mid =\mid \Delta _{R}\mid =\Delta $, $%
h_{L}=h_{R}=h $, $\mid \mathbf{Q}_{L}\mid =\mid \mathbf{Q}_{R}\mid =Q$ where
$\mathbf{Q}_{L,R}=\chi _{L,R}Q\mathbf{\hat{z}}$ are orthogonal to the
tunnelling barrier with the spiral helicity $\chi _{L(R)}=\pm 1$ for $%
\mathbf{Q}_{L,R}$ along (+) or opposite(-) to the z-axis. Note, that such a
junction is characterized by the \textit{superconducting phase} $\varphi
=\varphi _{L}-\varphi _{R}\neq 0$ and \textit{magnetic phase} $\theta
=\theta _{L}-\theta _{R}\neq 0$. It turns out that besides the singlet
amplitude $F_{\uparrow \downarrow }(F_{\downarrow \uparrow })$ the \textit{%
triplet pairing amplitudes} $F_{\uparrow \uparrow }$ $(F_{\downarrow
\downarrow })$ play very important role in the Josephson effect \cite{Kic-Ig}%
. (Note, that $F_{\uparrow \uparrow }$ $(F_{\downarrow \downarrow })$ was
first introduced in \cite{BuKuRu} in the study of the spiral magnetic order
in SC and first applied to the Josephson junction in \cite{Kic-Ig}. Later on
this effect was rediscovered by the Efetov's group in studying S-F-F-S
structures with rotating magnetization \cite{Efetov}, where $F_{\uparrow
\uparrow }$ $(F_{\downarrow \downarrow })$ give rise for new effects see
this issue.) It turns out that the Josephson current contains two parts

\begin{figure}[tbp]
\resizebox{.8 \textwidth}{!} {
\includegraphics*[width=8cm]{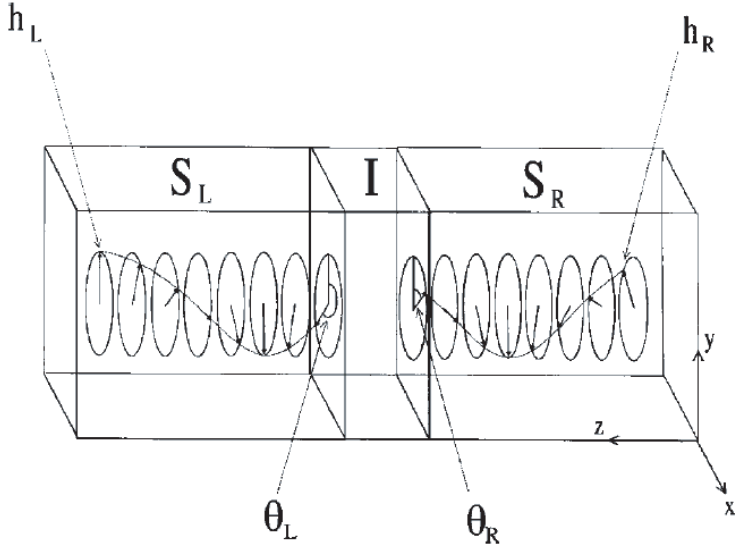}}
\caption{The Josephson junction with the insulating contact. $S_{L}$ and $%
S_{R}$ are superconductors with spiral magnetic order. The exchange fields $%
\vec{h}_{L,R}$ at the surface make angles $\protect\theta_{L,R}$with the
y-axis. $\vec{Q}_{L,R}$ are along the z-axis.}
\end{figure}

\begin{equation}
J_{J}(\varphi ,\theta )=(J_{c}^{s}-J_{-\chi }^{t}\cos \theta )\sin \varphi ,
\label{34}
\end{equation}%
where
\begin{equation}
J_{c}^{s}\sim T\sum_{\mathbf{k}_{L},\mathbf{k}_{R},\omega _{n}}\mid T_{%
\mathbf{k}_{L},\mathbf{k}_{R}}\mid ^{2}F_{\uparrow \downarrow }^{\dagger }(%
\mathbf{k}_{L},\omega _{n})F_{\uparrow \downarrow }(\mathbf{k}_{R},-\omega
_{n})  \label{35}
\end{equation}%
(and $J_{c}^{s}\sim \Delta ^{2}$) is due to the \textit{singlet amplitude }
and

\[
J_{-\chi }^t \sim -T\sum_{\mathbf{k}_{L},\mathbf{k}_{R},\omega _{n}}\mid T_{%
\mathbf{k}_{L},\mathbf{k}_{R}}\mid ^{2}\{F_{\uparrow \uparrow }^{\dagger }(%
\mathbf{k}_{L},\omega _{n})[F_{\uparrow \uparrow }^{\dagger }(\mathbf{k}%
_{R},-\omega _{n})]^{\ast }
\]%
\begin{equation}
+F_{\downarrow \downarrow }^{\dagger }(\mathbf{k}_{L},\omega
_{n})[F_{\downarrow \downarrow }^{\dagger }(\mathbf{k}_{R},-\omega
_{n})]^{\ast }\}  \label{36}
\end{equation}%
is due to the\textit{\ triplet amplitude}. It turns out that

\begin{equation}
J_{-\chi }^{t}\sim \Delta ^{2}h^{2}[f_{1}+(\chi _{L}\chi _{R})f_{2}(\Delta
,h)],  \label{37}
\end{equation}%
where $f_{1,2}(\Delta ,h)$ are given in \cite{Kic-Ig}, while $\chi =\chi
_{L}\chi _{R}$ is the total helicity (in \cite{Efetov} renamed to chirality)
of the junction. It was shown in \cite{Kic-Ig} that in some parameter region
the triplet effects dominate, i.e. $\mid J_{-\chi }^{t}\mid >J_{c}^{s}$,
thus giving rise to the $\pi $-Josephson junction. From Eq.(54) it is clear
that by changing the magnetic phase $\theta $ and chirality $\chi $ one can
tune the system from $0-$ to $\pi -$ junction. This new degree of freedom in
the junction - the magnetic phase $\theta $, first proposed in \cite{Kic-Ig}%
, opens a new possibility for \textit{switching elements} and \textit{%
quantum computing}. From the physical point of view the above model is a
\textit{paradigm} for analogous effects in S-F-F-S structures with rotating
magnetization, in which case is $\theta $ the angle between magnetizations
in neighbouring layers.

\subsection{Combined superconducting and magnetic Josephson effect}

In \cite{Kic-Ig-QC} the above model is developed further by including the
tunnelling of electronic spins and their effect on the energy of the
contact. Namely, in ferromagnetic superconductors with rotating of
magnetization (such as spiral) besides the standard Green's function $%
G_{\uparrow \uparrow }$ $(G_{\downarrow \downarrow })$, $F_{\uparrow
\downarrow }(F_{\downarrow \uparrow })$ for singlet SC other Green's
functions $G_{\uparrow \downarrow }$ $(G_{\downarrow \uparrow })$ and $%
F_{\uparrow \uparrow }$ $(F_{\downarrow \downarrow })$ are important \cite%
{Kic-Ig}, \cite{Kic-Ig-QC}, since they can produce a \textit{static spin
current} $J_{spin}$ through the junction (in absence of voltage),

\begin{equation}
J_{spin}=J_{spin,G}+J_{spin,F}  \label{38a}
\end{equation}%
where

\begin{equation}
J_{spin,G}\sim \sum \mid T\mid ^{2}(G_{\uparrow \downarrow ,L}G_{\downarrow
\uparrow ,R}-G_{\downarrow \uparrow ,L}G_{\uparrow \downarrow ,R}]\sim
h^{2}\sin \theta .  \label{38}
\end{equation}

\[
J_{spin,F}\sim \sum \mid T\mid ^{2}[F_{\uparrow \uparrow }^{\dagger }(%
\mathbf{k}_{L},\omega _{n})[F_{\uparrow \uparrow }^{\dagger }(\mathbf{k}%
_{R},-\omega _{n})]^{\ast }
\]

\begin{equation}
-F_{\downarrow \downarrow }^{\dagger }(\mathbf{k}_{L},\omega
_{n})[F_{\downarrow \downarrow }^{\dagger }(\mathbf{k}_{R},-\omega
_{n})]^{\ast }\}\sim h^{2}\Delta ^{2}\tilde{f}_{\chi }\cos \varphi \sin
\theta .  \label{38b}
\end{equation}%
The exact expression for $J_{spin,G}$ and $J_{spin,F}$ will be published
elsewhere \cite{Kic-Ig-QC}. The energy of this combined \textit{magnetic and
superconducting Josephson junction }$E=E_{mJ}(\theta )+E_{J}(\varphi ,\theta
)$ is

\begin{equation}
E(\theta ,\varphi )=-Ah^{2}\cos \theta -\Delta ^{2}(B+C_{\chi }h^{2}\cos
\theta )\cos \varphi ,  \label{39}
\end{equation}%
where the cumbersome expressions for $A,B,C$ are given in \cite{Kic-Ig-QC}.
Note, that both the spin $J_{spin}(\theta ,\varphi )\sim \partial E/\partial
\theta $ and the charge $J_{J}(\varphi ,\theta )\sim \partial E/\partial
\varphi $ Josephson current depend on $\varphi $ and $\theta $. Thus, by
tuning $\theta $ and $\varphi $ one can tune these currents. In case of
small contacts with small charge and "spin" capacitance the system is in the%
\textit{\ quantum regime} thus giving possibility for a novel Josephson
qubit. In fact the latter consists from \textit{two qubits} - the
superconducting and magnetic one, which is of a potential interest for
applications \cite{Kic-Ig-QC}.

\section{Conclusion}

The rare earth ternary compounds are reach physical systems which allow
coexistence of superconductivity and various magnetic orders, such as
ferromagnetic, antiferromagnetic, weak ferromagnetism. It turns out that in
these systems superconductivity and ferromagnetism never coexist and the
latter is modified into a spiral or domain structure - depending on magnetic
anisotropy. This is realized in rare earths ternary compounds as well as in $%
AuIn_{2}$ where the modified nuclear ferromagnetism and superconductivity
coexist. Although the antiferromagnetic order and superconductivity coexist
much easier, these systems show peculiar behavior in the presence of
nonmagnetic impurities which surprisingly act as pair-breakers. In case when
the antiferromagnetic order is accompanied by the weak ferromagnetism new
coexistence phases appear - the Meissner and spontaneous vortex state.
Magnetic superconductors show peculiar behavior in magnetic field. Near the
magnetic critical temperature the upper critical field goes to zero faster
than the thermodynamical field, thus giving rise to the first order
transition. Various phases are possible in the $H-T$ diagram depending on
the purity and demagnetization effects of real samples. The lower critical
field is weakly affected by the exchange field due to localized moments.

The Josephson junctions based on bulk ferromagnetic superconductors with
spiral order are characterized by the superconducting and magnetic phase,
opening possibilities for a new kind of coupled qubits. The triplet pairing
amplitude gives rise to the $\pi -$junction which can be tuned by changing
the magnetic phase and chirality.

This article is submitted for the \textit{Special Issue Comptes de
l'Academie des Sciences: Problems of the Coexistence of Magnetism and
Superconductivity} edited by A. Buzdin.

\textit{Acknowledgments} - I would like to thank Alexander Buzdin for
collaboration, support and hospitality I have enjoyed during my stay, in
summer 2005, at the University of Bordeaux when this review was written. I
thank Igor Kuli\'{c} for collaboration and support.

\section{Appendix}

The problem of the coexistence of SC and magnetic order with a wave vector Q
can be studied in principle by using Gorkov equations for any Q. However, in
systems where $a<<Q^{-1}$ is fulfilled the quasiclassical
Eilenberger-Larkin-Ovchinikov (ELO) equations are more suitable and
efficient.

\subsection{Gorkov equations for MS}

These equations contain normal and anomalous Green's functions $(\hat{G}%
)_{\alpha \beta }(x,y)=-<\hat{T}\psi _{\alpha }(x)\psi _{\beta }^{\dagger
}(y)>$ and $(\hat{F}^{\dagger })_{\alpha \beta }(x,y)=<\hat{T}\psi _{\alpha
}^{\dagger }(x)\psi _{\beta }^{\dagger }(y)>$, where $x\equiv (\vec{r},\tau
) $ and $\alpha ,\beta =\uparrow ,\downarrow $. The superconducting order
parameter is defined by

\begin{equation}
\Delta ^{\ast }(\vec{r}_{i}-\vec{r}_{j})=-\frac{1}{2}V(\vec{r}_{i}-\vec{r}%
_{j})[<\psi _{\uparrow }^{\dagger }(\vec{r}_{i})\psi _{\downarrow }^{\dagger
}(\vec{r}_{j})>-<\psi _{\downarrow }^{\dagger }(\vec{r}_{i})\psi _{\uparrow
}^{\dagger }(\vec{r}_{j})>]  \label{40}
\end{equation}
where the pairing interaction $V(\vec{r}_{i}-\vec{r}_{j})$ is responsible
for the $s-wave$ pairing in absence of magnetic order. Note, that in the
presence of any (in)commensurate magnetic order the superconducting order
parameter should be nonuniform, i.e. $\Delta (\vec{r}_{1},\vec{r}%
_{2})=\Delta (\vec{r}_{1}-\vec{r}_{2},\vec{r}),$ $\vec{r}=(\vec{r}_{i}+\vec{r%
}_{j})/2,$ However, it has been shown in\cite{BuKuRu} that when $h<\hbar
v_{F}Q$ the nonuniform part of $\Delta$ is small and of the order $(h/\hbar
v_{F}Q)\Delta $. That is the reason for our assumption $\Delta ^{\ast }(\vec{%
r}_{i},\vec{r}_{j})=\Delta ^{\ast }(\vec{r}_{i}-\vec{r}_{j}).$ The set of
equations for the Green's functions is given by

\[
\lbrack i\omega _{n}-\hat{\epsilon}_{0}(\hat{p})-\hat{V}_{ex}(\vec{r})]\hat{G%
}_{\omega _{n}}(\vec{r},\vec{r}^{\prime })+\int d^{2}x\hat{\Delta}(\vec{r}-%
\vec{x})\hat{F}_{\omega _{n}}^{\dagger }(\vec{x},\vec{r}^{\prime })=\delta (%
\vec{r}-\vec{r}^{\prime }),
\]%
\begin{equation}
\lbrack i\omega _{n}+\hat{\epsilon}_{0}(\hat{p})+\hat{V}_{ex}^{tr}(\vec{r})]%
\hat{F}_{\omega _{n}}^{\dagger }(\vec{r},\vec{r}^{\prime })-\int d^{2}x\hat{%
\Delta}^{\ast }(\vec{r}-\vec{x})\hat{G}_{\omega _{n}}(\vec{x},\vec{r}%
^{\prime })=0,  \label{41}
\end{equation}
where
\begin{equation}
\hat{\Delta}=\Delta \left[
\begin{array}{cc}
0 & 1 \\
-1 & 0%
\end{array}%
\right] ,\hat{V}_{ex}(\vec{r})=\left[
\begin{array}{cc}
h^{z}(\mathbf{r}) & h^{\perp }e^{-i(\mathbf{Qr+}\theta \mathbf{)}} \\
h^{\perp }e^{i(\mathbf{Qr+}\theta )} & -h^{z}(\mathbf{r})%
\end{array}%
\right] .  \label{42}
\end{equation}

In case of a spiral magnetic ordering (with spatial rotation of
magnetization) besides the singlet pairing amplitude $F_{\downarrow \uparrow
}^{\dagger }\sim \Delta $ there is also a \textit{triplet pairing amplitude }%
$F_{\uparrow \uparrow }^{\dagger }\sim \Delta \cdot h_{\theta }^{\ast }$
(and $F_{\downarrow \downarrow }^{\dagger }$) \cite{BuKuRu}. $F_{\downarrow
\uparrow }^{\dagger }$ is responsible for \textit{singlet pairing} with the
order parameter $\Delta _{\uparrow \downarrow }(\mathbf{r})$, while the
triplet amplitude $F_{\uparrow \uparrow }^{\dagger }$, which is due to the
rotating magnetization which mixes spin up and down, does not give rise to
the triplet pairing since it is assumed from the very beginning that $\Delta
_{\uparrow \uparrow }=\Delta _{\downarrow \downarrow }=0$. Note that $%
F_{\uparrow \uparrow }^{\dagger }$ contains two terms - the first one is
\textit{odd in frequency} $\omega _{n}$ and the second one is \textit{odd in
momentum} $\mathbf{k}$.

\subsection{ELO equations for MS}

In problems related to the presence of the exchange field $\mathbf{h(r})=h(%
\mathbf{r})\mathbf{e}_{z}$ (oriented along the $z$-axis) and of nonmagnetic
potential and spin-orbit scattering a generalization of the $ELO$ equations
is needed \cite{BuBuKuPa}, \cite{Demler}. In that case the Gor'kov equations
contain $4\times 4$ Green's functions $\hat{G}(x_{1},x_{2})=-\langle \hat{T}%
\hat{\Psi}(x_{1})\hat{\Psi}^{\dagger }(x_{2})\rangle $ with the
four-component spinor $\hat{\Psi}^{\dagger }(x_{1})=(\hat{\psi}_{\uparrow
}^{\dagger }\hat{\psi}_{\downarrow }^{\dagger }\psi _{\uparrow }\psi
_{\downarrow })$. Here $\hat{\tau}_{i}$ are Nambu matrices in the
electron-hole space and $\hat{\sigma}_{i}$ are Pauli matrices in the
spin-space and $x\equiv (\mathbf{r},\tau )$. (Note, that here the functions $%
F$ are defined with minus sign.) The SC order parameter is $\hat{\Delta}(%
\mathbf{r})=i\hat{\tau}_{+}\hat{\sigma}_{2}\Delta (\mathbf{r})-i\hat{\tau}%
_{-}\hat{\sigma}_{2}\Delta ^{\ast }(\mathbf{r})$ with

\begin{equation}
\Delta (\mathbf{r})=iVTr\{\hat{\tau}_{+}\hat{\sigma}_{2}\hat{G}(x,x)\}/2.
\label{43}
\end{equation}
By integrating out unimportant degrees of freedom at small distances one
defines the quasiclassical Green's function $\hat{g}_{\omega _{n}}(\mathbf{R}%
,\mathbf{p}_{F})$

\begin{equation}
\hat{g}_{\omega _{n}}(\mathbf{R},\mathbf{p}_{F})=\int d\xi _{p}\int d(%
\mathbf{r}_{1}-\mathbf{r}_{2})e^{-i\mathbf{p}((\mathbf{r}_{1}-\mathbf{r}%
_{2})}\hat{\tau}_{3}\hat{G}(\mathbf{r}_{1},\mathbf{r}_{2},\omega _{n}).
\label{44}
\end{equation}
Then the ELO equations for the quasiclassical Green's functions read

\begin{equation}
i\mathbf{v}_{F}\nabla _{\mathbf{R}}\hat{g}_{\omega _{n}}=[i\omega _{n}\hat{%
\tau}_{3}+\hat{\tau}_{3}\hat{\Delta}(\mathbf{R})+h(\mathbf{R})\hat{\sigma}%
_{3}-\hat{\tau}_{3}\Sigma _{\omega _{n}}^{imp}(\mathbf{R,p}_{F}),\hat{g}%
_{\omega _{n}}],  \label{45}
\end{equation}
where the impurity self-energy in the Born approximation is given by

\begin{equation}
\Sigma _{\omega _{n}}^{imp}(\mathbf{R,p}_{F})=c_{i}N(0)\langle \hat{U}(%
\mathbf{p}_{F}-\mathbf{p}_{F}^{\prime })\hat{g}_{\omega _{n}}(\mathbf{R,p}%
_{F}^{\prime })\hat{U}(\mathbf{p}_{F}-\mathbf{p}_{F}^{\prime })\rangle _{%
\mathbf{p}_{F}^{\prime }}.  \label{46}
\end{equation}
The impurity potential (matrix) contains the non-magnetic and spin-orbit
scattering

\begin{equation}
\hat{U}(\mathbf{p}_{F}-\mathbf{p}_{F}^{\prime })=U_{1}\hat{\tau}_{3}+iU_{so}[%
\mathbf{\hat{p}}_{F}\times \mathbf{\hat{p}}_{F}^{\prime }]\mathbf{\hat{\alpha%
},}  \label{47}
\end{equation}
with the vector matrix $\mathbf{\hat{\alpha}}=[(1+\hat{\tau}_{3})\mathbf{%
\sigma }+(1-\hat{\tau}_{3})\sigma _{2}\mathbf{\sigma }\sigma _{2}]/2$ and $%
\mathbf{\hat{p}}_{F}=\mathbf{p}_{F}/\mid \mathbf{p}_{F}\mid $. The matrix $%
\hat{g}_{\omega _{n}}(\mathbf{R,p}_{F})$ is given by
\begin{equation}
\hat{g}_{\omega _{n}}(\mathbf{R,p}_{F})=-i\left(
\begin{array}{cccc}
g_{+} & 0 & 0 & if_{+} \\
0 & g_{-} & if_{-} & o \\
0 & -if_{-}^{\dagger } & -g_{-} & 0 \\
-if_{+}^{\dagger } & 0 & 0 & -g_{+}%
\end{array}%
\right) .  \label{48}
\end{equation}
By including also orbital effects in magnetic field, which is determined by
the vector potential $\mathbf{A(R)}$, the $ELO$ equations written in
components $g_{\pm }$ and $f_{\pm }$ read

\[
\lbrack \tilde{\omega}_{n,\pm }+ie\mathbf{v}_{F}\cdot \mathbf{A(R)}\pm ih(%
\mathbf{R})-\frac{1}{2}\mathbf{v}_{F}\cdot \nabla _{\mathbf{R}}]f_{\pm }(%
\mathbf{p}_{F},\mathbf{R},\omega _{n})
\]

\begin{equation}
=\tilde{\Delta}_{\pm }(\mathbf{p}_{F},\mathbf{R},\omega _{n})g_{\pm }(%
\mathbf{p}_{F},\mathbf{R},\omega _{n}),  \label{49}
\end{equation}

\[
\lbrack \tilde{\omega}_{n,\pm }+ie\mathbf{v}_{F}\cdot \mathbf{A(R)}\pm ih(%
\mathbf{R})+\frac{1}{2}\mathbf{v}_{F}\cdot \nabla _{\mathbf{R}}]f_{\pm
}^{\dagger }(\mathbf{p}_{F},\mathbf{R},\omega _{n})
\]

\begin{equation}
=\Delta _{\pm }^{\ast }(\mathbf{p}_{F},\mathbf{R,}\omega _{n})g_{\pm }(%
\mathbf{p}_{F},\mathbf{R},\omega _{n}),  \label{50}
\end{equation}
with the normalization condition and the self-consistency equation,
respectively
\begin{equation}
g_{\pm }^{2}+f_{\pm }^{\dagger }f_{\pm }=1  \label{51}
\end{equation}
\begin{equation}
\Delta (\mathbf{p}_{F},\mathbf{R})=\frac{\pi T}{2}\sum_{\omega _{n},\mathbf{p%
}_{F}^{\prime }}V(\mathbf{p}_{F},\mathbf{p}_{F}^{\prime })\{f_{+}(\mathbf{R},%
\mathbf{p}_{F}^{\prime },\omega _{n})+f_{-}(\mathbf{R},\mathbf{p}%
_{F}^{\prime },\omega _{n})\}.  \label{52}
\end{equation}
$\tilde{\omega}_{n,\pm }$ and $\tilde{\Delta}_{\pm }(\mathbf{p}_{F},\mathbf{R%
},\omega _{n})$ are defined by

\[
\tilde{\omega}_{n,\pm }=\omega _{n}+\frac{1}{2\tau _{1}}\langle g_{\pm }(%
\mathbf{p}_{F}^{\prime },\mathbf{R},\omega _{n})\rangle _{\mathbf{p}%
_{F}^{\prime }}
\]

\begin{equation}
+\frac{3}{2\tau _{so}}\langle g_{\mp }(\mathbf{p}_{F}^{\prime },\mathbf{R}%
,\omega _{n})\sin ^{2}(\theta -\theta ^{\prime })\rangle _{\mathbf{p}%
_{F}^{\prime }},  \label{53}
\end{equation}

\[
\tilde{\Delta}_{\pm }(\mathbf{p}_{F},\mathbf{R},\omega _{n})=\Delta (\mathbf{%
p}_{F},\mathbf{R})+\frac{\Gamma _{1}}{2}\langle f_{\pm }(\mathbf{p}%
_{F}^{\prime },\mathbf{R},\omega _{n})\rangle _{\mathbf{p}_{F}^{\prime }}
\]

\begin{equation}
+\Gamma _{so}\langle f_{\mp }(\mathbf{p}_{F}^{\prime },\mathbf{R},\omega
_{n})\sin ^{2}(\theta -\theta ^{\prime })\rangle _{\mathbf{p}_{F}^{\prime }}
\label{54}
\end{equation}
where $\Gamma _{1}=c_{i}N(0)\mid U_{1}\mid ^{2}$ and $\Gamma
_{so}=c_{i}N(0)\mid U_{so}\mid ^{2}$.

The microscopic theory of magnetic superconductors which takes into account
spin (exchange) and orbital (electromagnetic) effects of magnetic order, as
well as of nonmagnetic impurity scattering, has been developed by using the
above generalized $ELO$ equations \cite{BuBuKuPaAdvances}, \cite{BuBuKuPa}.

\end{document}